\documentclass[a4paper,12pt]{article}
\usepackage{jheppub,esint,shuffle,psfrag}
\usepackage[utf8]{inputenc}

\usepackage{epsfig,amssymb,amsmath,psfrag,subfigure,rotate,color,wasysym,xcolor}
\usepackage[bbgreekl]{mathbbol}

\usepackage{booktabs}

\allowdisplaybreaks
\newcommand{\insertfig}[2]{\includegraphics[width=#1cm]{#2}}

\DeclareSymbolFontAlphabet{\mathbbm}{bbold}
\DeclareSymbolFontAlphabet{\mathbb}{AMSb}%

\def \lab #1 {\label{#1}}

\newcommand\re[1]{(\ref{#1})}
\def\d{\hbox{{d}\kern-.20em\hbox{l}}}
\def \qqquad {\qquad\quad}
\def \qqqquad {\qquad\qquad}
\def \matrix #1 {\left(\begin{array}{cc} #1 \end{array}\right)}

\def \tr {\mathop{\rm tr}\nolimits}

\def \e  {\mathop{\rm e}\nolimits}
\newcommand\lr[1]{{\left({#1}\right)}}

\newcommand \vev [1] {\langle{#1}\rangle}

\def\1{\hbox{{1}\kern-.25em\hbox{l}}}

\newcommand{\ft}[2]{{\textstyle\frac{#1}{#2}}}

\title{ 
Crossing bridges with strong Szeg\H{o} limit theorem
}
 
\author[a,b]{A.V.~Belitsky}
\author [b]{and G.P.~Korchemsky}
 \affiliation[a] {Department of Physics, Arizona State University, 
 Tempe, AZ 85287-1504, USA}

 \affiliation[b] {Institut de Physique Th\'eorique\footnote{Unit\'e Mixte de Recherche 3681 du CNRS}, Universit\'e Paris Saclay, CNRS, CEA, 91191 Gif-sur-Yvette} 
 
\preprint{  \parbox[t]{28mm}{IPhT--T20/038}}

 \abstract
{
We develop a new technique for computing a class of four-point correlation functions of heavy half-BPS operators in planar $\mathcal N=4$ SYM theory which admit factorization into a product 
of two octagon form factors with an arbitrary bridge length. We show that the octagon can be expressed as the Fredholm determinant of the integrable Bessel operator and demonstrate that this 
representation is very efficient in finding the octagons both at weak and strong coupling.  At weak coupling, in the limit when the four half-BPS operators become null separated in a sequential 
manner, the octagon obeys the Toda lattice equations and can be found in a closed form. At strong coupling, we exploit the strong Szeg\H{o} limit theorem to derive the leading asymptotic behavior 
of the octagon and, then, apply the method of differential equations to determine the remaining subleading terms of the strong coupling expansion to any order in the inverse coupling. To achieve 
this goal, we generalize results available in the literature for the asymptotic behavior of the determinant of the Bessel operator. As a byproduct of our analysis, we formulate a 
Szeg\H{o}-Akhiezer-Kac formula for the determinant of the Bessel operator with a Fisher-Hartwig singularity and develop a systematic approach to account for subleading power suppressed contributions.
 }
 
\begin{document}

\maketitle
\flushbottom
\setcounter{footnote} 0

\section{Introduction}

This paper is devoted to the study of four-point correlation functions of half-BPS single-trace operators in four-dimensional maximally supersymmetric Yang-Mills theory ($\mathcal N=4$ SYM). The 
theory is integrable in the planar limit~\cite{Beisert:2010jr} and it is believed that various observables can be computed exactly for an arbitrary value of 't Hooft coupling constant. 

In general, the four-point correlation functions of half-BPS operators are complicated observables  in $\mathcal N=4$ SYM and their explicit expressions are available for the first few orders of the weak 
coupling expansion, at best. At strong coupling, the AdS/CFT correspondence predicts that, in the planar limit, the correlators coincide with scattering amplitudes of four closed string states dual to the
respective four half-BPS operators located at the boundary of AdS. Their calculation requires quantization of strings on $AdS_5\times S^5$ target space and can not be performed by currently existing 
methods.

\begin{figure}[h!t]
\centerline{\insertfig{17}{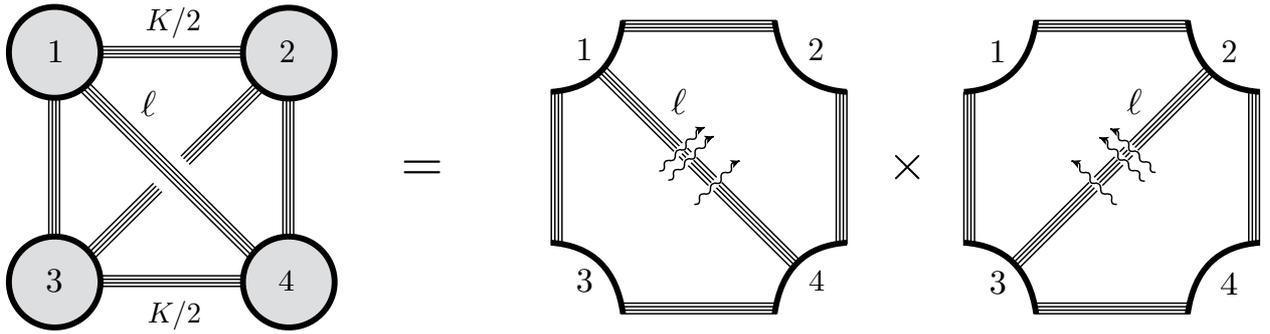}}
\caption{Diagrammatical representation of the four-point correlation function of heavy half-BPS operators in the Born approximation
(left-hand side of the equality) and at finite 't Hooft coupling (right-hand side). The half-BPS operators are represented by 
the four grey blobs with exactly $K$ outgoing lines, solid lines denote scalar propagators with the accompanying label counting their total number. The `simplest' correlator corresponds to the zero length bridge 
$\ell = 0$, the `asymptotic' one is given by the sum over $\ell$ ranging from $0$ to $K/2$. Factorization of the correlator into the product of two octagons is shown in the right-hand side of the equation. Each 
octagon describes excitations (wave lines) propagating on a world-sheet of an open string and crossing the bridge of length $\ell$.}\label{fig:frames}
\end{figure}

It was recently recognized that difficulties at weak and strong coupling can be alleviated by a judicial choice of the half-BPS operators~\cite{Coronado:2018ypq}. The latter are built from $K$ scalar fields 
(with $K\ge 2$) each carrying one unit of the $R-$charge. Their four-point correlation function at Born level (zero coupling constant) is given by the product of free scalar propagators stretched between the 
four operators in a pairwise manner (see the leftmost panel in Figure~\ref{fig:frames}). These bunches are called \textit{bridges}. The number of propagators in each of them defines the length (or rather width) 
of the connecting bridge. A generic contributing graph 
contains six bridges whose length varies between $0$ and $K$. Choosing the half-BPS operators appropriately and taking their $R-$charge $K$ to be arbitrarily large, one can ensure that the four bridges 
connecting the four operators in a sequential manner have a large length $O(K)$ whereas the length of the remaining two stays finite.

Two types of such correlation functions dubbed the {\it simplest} and {\it asymptotic} in Ref.~\cite{Coronado:2018ypq} are shown schematically in Figure~\ref{fig:frames}. Their distinguished feature is 
that, for finite 't Hooft coupling in planar $\mathcal N=$ SYM, they can be expressed in terms of new building blocks $\mathbb O_\ell$ dubbed the \textit{octagons}. 
The correlation functions in question can be viewed as two polygons glued together along the four bridges of large lengths. The octagon describes each of these polygons
and depends on a nonvanishing length $\ell$ of the internal bridge. In the limit when the lengths of all glued-up bridges around the polygon perimeter are large, the four-point functions factorize 
into  a bi-linear combination of octagons.

\medskip

The definition of the octagon relies on a dual description of the correlation functions in planar $\mathcal N=4$ SYM in terms of an effective two-dimensional integrable theory describing the string world-sheet in the 
AdS/CFT correspondence~\cite{Basso:2015zoa,Fleury:2016ykk,Eden:2016xvg,Bajnok:2017mdf}. The octagon takes into account the propagation of excitations (known as magnons) with the (mirror) energy $E$ on 
the world-sheet across the internal bridge $\ell$. Their contribution to $\mathbb  O_\ell$ is 
proportional to $\e^{-E\ell}$ and is accompanied by the product of two hexagon form factors encoding magnon scattering. This leads to a representation of the octagon 
as an infinite sum over excitations crossing the bridge of finite length $\ell$~\cite{Coronado:2018ypq}. It is similar to analogous form factor representations of two-point correlation functions of 
local operators separated by a proper time 
$\ell$ in integrable models, see, e.g., Ref.~\cite{Korepin:1993kvr}.

The hexagon form factors are known for arbitrary 't Hooft coupling \cite{Basso:2015zoa} as a solution to bootstrap form factor axioms \cite{Smirnov:1992vz}. Thus, having evaluated the aforementioned sums 
over all excitations, one would obtain a finite-coupling representation of the octagons. This was accomplished in Refs.\ \cite{Kostov:2019stn,Kostov:2019auq}, where a concise formula for $\mathbb  O_{\ell}$ 
was given in terms of a determinant of a semi-infinite matrix. This result served as the starting point of our analysis in Refs.\ \cite{Belitsky:2019fan,Belitsky:2020qrm}, where the octagon at zero-length bridge $\ell=0$ 
was further cast as the Fredholm determinant of an integral operator acting on a semi-infinite line. The kernel of the operator in question turned out to be closely related to the Bessel kernel that previously appeared in the 
study of the Laguerre ensemble in random matrix theory~\cite{Forrester:1993vtx,Tracy:1993xj}. Taking advantage of this property, we constructed in Ref.~\cite{Belitsky:2020qrm} a systematic expansion of the octagon 
$\mathbb  O_{\ell=0}$ at strong coupling by applying the method of differential equations originally developed in Refs.~\cite{Its:1990,Korepin:1993kvr} for calculation of two-point correlation functions in integrable models. 

Strong coupling expansion of the octagon $\mathbb O_\ell$ takes the following general form
\begin{align}\label{O0}
\mathbb  O_{\ell} = \exp\lr{ - g A_0 + \frac{1}{2}A_1^2 \log g + B + \frac{A_2}{2 g} + O(1/g^2)}\,,
\end{align}
where the expansion coefficients $A_k$ and $B$ (with $k=0,1,2,\dots$) depend on the bridge length $\ell$ as well as on the coordinates of the four operators. The coefficient $B$ defines the coupling-independent 
correction to the exponent in \re{O0} and it plays a special role in the method of differential equations. It arises there as an arbitrary integration constant and, in order to find it, we have to rely on another approach. 

Our main goal in the present work is to extend the findings of Refs.\ \cite{Belitsky:2019fan,Belitsky:2020qrm} to the octagon $\mathbb O_{\ell}$ with an arbitrary bridge length $\ell$ and to determine the missing 
coefficient function $B$. We demonstrate that this can be achieved by combining the method of differential equations with another powerful technique based on the strong Szeg\H{o} limit 
theorem~\cite{Szego:1915,Szego:1952}. 

This theorem describes the asymptotic behavior of determinants of Toeplitz matrices of the form $T_n(a)=(a_{j-k})_{j,k=0}^n$, with $a_k$ being the Fourier coefficients 
of a function $a(\e^{i\theta})$ defined on a unit circle, in the limit when their size goes to infinity, $n\to\infty$. For sufficiently smooth functions $a$, it reads
\begin{align}\label{th}
 \det T_n (a) = \exp\lr{-n \tilde A_0 + \tilde B+\dots}
\, ,
\end{align}
where the first two terms are given by $\tilde A_0=-(\log a)_0$ and $\tilde B= \sum_{k=1}^\infty k (\log a)_k (\log a)_{- k}$, with $(\log a)_k$ being 
the Fourier coefficients of the function $\log (a(\e^{i\theta}))$, and the ellipses denote contributions vanishing as $n\to\infty$. A comprehensive review of the strong Szeg\H{o} limit theorem can be found in Refs.~\cite{Bttcher2006AnalysisOT,Basor:2012,Krasovsky13}.

We observe that for $n\sim g$, the expressions on the right-hand side of \re{O0} and \re{th} possess similar forms. The reason for this is that, as we argue below, the relation \re{O0} can be derived by applying 
the strong Szeg\H{o} limit theorem to a certain integral operator. Notice, however, that in counter-distinction to \re{th}, the relation \re{O0} contains a logarithmically enhanced term. We explain its origin in a moment.

We should mention that this is not the first time that the strong Szeg\H{o} limit theorem comes to the rescue in solving complicated physical problems. A notable example is a calculation of the correlation function 
$M_n^2 = \vev{\sigma_{0,0} \, \sigma_{0,n}}$ of two spins separated by $n$ lattice sites  in the two-dimensional Ising model (see Refs.\ \cite{Bottcher:1995,Bottcher:1999,Basor:2012,Krasovsky13}, for a historical 
review). This correlation function can be cast into the form of the Toeplitz determinant $\det T_n(a)$ with the symbol $a$ depending on parameters of the model. The strong Szeg\H{o} limit theorem predicts the 
asymptotic behavior of $M_n$ at large $n$ which coincides with the celebrated Onsager formula for the magnetization in the Ising model.

Analyzing the octagons in $\mathcal{N} = 4$ SYM theory, we encounter a continuous version of Toeplitz determinants. The generalization of the strong Szeg\H{o} limit theorem to this case was devised by 
Akhiezer \cite{Akhiezer:1964} and Kac \cite{Kac:1964} and it is known as the Szeg\H{o}-Akhiezer-Kac formula. This formula provides the asymptotic behavior of determinants of integral operators with sufficiently 
smooth kernels acting on a finite interval $[0,R]$ as $R \to \infty$. It takes the form similar to \re{O0} with $R\sim g$ and yields a definite prediction for the first three terms in the exponent of \re{O0}.

We show below that, for an arbitrary bridge length $\ell$, the octagon $\mathbb O_\ell$ is given by the Fredholm determinant of the so-called truncated Bessel operator acting on the interval $[0,2g]$. Applying the 
Szeg\H{o}-Akhiezer-Kac formula at strong coupling, i.e., for $g\gg 1$, we recover the known expression for the leading $O(g)$ term in \re{O0} but obtain a divergent result for the coefficient function $B$. This 
happens because the integral kernel of the Bessel operator is singular and it does not fulfill the applicability conditions of the strong Szeg\H{o} limit theorem. We identify the corresponding singularity as a  
Fisher-Hartwig singularity~\cite{Fisher68}. Its presence modifies the asymptotic behavior of the octagon and generates an additional $\log g$ contribution in \re{O0}. 

At present, generalizations of the Szeg\H{o}-Akhiezer-Kac formula to the Bessel operator with Fisher-Hartwig singularities is only known for specific (unphysical) values $\ell=\pm 1/2$, see Ref.~\cite{BasorEhrhardt05}. For our purposes, we 
have to extend these results to arbitrary $\ell$. We do it by formulating a conjecture for the determinant of the Bessel operator with a  Fisher-Hartwig singularity. Being applied to the octagon $\mathbb O_\ell$, 
it unambiguously fixes the coefficients $A_0$, $A_1$ and $B$ in \re{O0} for arbitrary $\ell$.~\footnote{The resulting expression for $B$ is conventionally called the Dyson-Widom constant.}
We verify that their values are in perfect agreement with outcomes of numerical computations. Taking thus obtained expressions for $A_0$, $A_1$ and $B$ as initial conditions, we apply the method 
of differential equations to determine all remaining coefficients in the strong coupling expansion of the octagon.

Our subsequent presentation is organized as follows. In Section~\ref{sect:corr}, we define four-point correlation functions of half-BPS operators and discuss their relation to the octagons in the limit when the 
operators become infinitely heavy. In Section~\ref{sect:oct}, we work out a representation of the octagon as the Fredholm determinant of the Bessel operator. We use this representation in Secton~\ref{sect:weak} 
to show that, at weak coupling, the octagon is given to any order in 't Hooft coupling by multilinear combinations of the so-called ladder integrals. We also demonstrate that in the null limit, when the four half-BPS operators 
sit at the vertices of a null rectangle, the octagon satisfies the Toda lattice equations and it can be found in a closed form. In Section~\ref{sect:str}, we formulate a conjectured generalization of the Szeg\H{o}-Akhiezer-Kac 
formula for the Bessel operator with a  Fisher-Hartwig singularity and apply it to determine the first three coefficients in the strong coupling expansion of the octagon \re{O0}. In Section~\ref{sect:exp}, we 
employ the method of differential equations to find the remaining expansion coefficients in \re{O0}. Properties of the resulting series in the inverse coupling are discussed in Section~\ref{sect:prop}. We 
demonstrate that certain class of corrections can be resummed to all orders, thus improving convergence of the strong coupling expansion. We compare the obtained expressions with numerical results and 
observe a perfect agreement. Concluding remarks are given in Section~\ref{sect:conc}. The Szeg\H{o}-Akhiezer-Kac formula for the Bessel operators with a  Fisher-Hartwig singularity is derived in 
Appendix~\ref{app:HF}. In Appendix~\ref{app:B}, we apply this formula to compute the Dyson-Widom constant $B$ in \re{O0} in different  kinematical regions and compare it with numerical results of 
Ref.~\cite{Belitsky:2020qrm}. In Appendix~\ref{app:WH}, we discuss a relation between the Fredholm determinants of Wiener-Hopf and Bessel operators. In Appendix~\ref{app:num}, we test the obtained expressions 
for the octagon by comparing them with the results of numerical computations. In Appendix \ref{AppendixClustering}, we obtain 
an equivalent representation for the leading term in the strong coupling expansion \re{O0}.   Finally, in Appendix \ref{app:sim} we construct the similarity transformation that allows us to simplify a determinant representation of the octagon.

\section{Correlation functions of heavy half-BPS operators}\label{sect:corr}

In this paper, we study four-point correlation functions of single-trace scalar half-BPS operators $\mathcal O_{K}= \tr[( y \Phi)^K]$ in planar $\mathcal N=4$ SYM. These operators are built out of  
$K$ copies of the scalar fields $(y\Phi)= \sum_{I=1}^6 y^I \Phi^I(x)$ projected onto  auxiliary six-dimensional null vectors $y^I$. The pair of variables $x^\mu$ and $y^I$ defines the coordinates 
of $\mathcal O_{K}$ in the space-time and isotopic $R-$space, respectively.

The scaling dimension of $\mathcal O_{K}$ is protected from quantum corrections and equals its $R-$charge, $\Delta_K=K$. In a similar manner, two- and three-point correlation functions of these operators 
do not depend on the 't Hooft coupling $g^2=g_{\rm YM}^2 N_c/(4 \pi)^2$ and coincide with their Born level expressions. Their four-point correlation function takes the following general form
\begin{align}\label{G4}
G_4=\vev{\mathcal O_K(1) \mathcal O_K(2) \mathcal O_K(3) \mathcal O_K(4)} = \lr{y_{12}^2 y_{34}^2\over x_{12}^2 x_{34}^2}^K
\mathcal G_K(z,\bar z,\alpha,\bar\alpha)\,,
\end{align}
where $\mathcal O_{K}(i)$ denotes the half-BPS operator with the coordinates $(x_i,y_i)$ and $\mathcal G(z,\bar z,\alpha,\bar\alpha)$ is a nontrivial function of the 't Hooft coupling. It also depends on two 
pairs of complex variables $(z,\bar z)$ and $(\alpha,\bar \alpha)$ defined by the two types of the cross-ratios 
\begin{align}\notag\label{x-ratio}
& z\bar z= {x_{12}^2 x_{34}^2\over x_{13}^2x_{24}^2} \,,\qqqquad
(1-z)(1-\bar z)= {x_{23}^2 x_{14}^2\over x_{13}^2x_{24}^2} \,,
\\
& \alpha\bar \alpha= {y_{12}^2 y_{34}^2\over y_{13}^2y_{24}^2} \,, \qqqquad
(1-\alpha)(1-\bar \alpha)= {y_{23}^2 y_{14}^2\over y_{13}^2y_{24}^2} \,,
\end{align}
where $x_{ij}\equiv (x_i-x_j)^2$ and $y_{ij}^2\equiv(y_i-y_j)^2$ with $y_i^2=0$. The variables $\alpha$ and $\bar\alpha$ are complex conjugate to each other. For variables $z$ and $\bar z$, the same 
holds true in the Euclidean signature, whereas in the Lorentzian signature they are independent. In what follows, we study the properties of the correlation function \re{G4} in the limit $K\to\infty$ when 
all four half-BPS operators become infinitely heavy.~\footnote{In planar $\mathcal N=4$ SYM, this limit corresponds to taking  $N_c\to\infty$ with $g^2$ fixed and, then, sending $K$ to infinity.}

It was previously realized that the correlation function \re{G4} reveals some interesting properties in the above limit. At weak coupling, explicit expressions for $G_4$ are known up to three loops 
for arbitrary weights $K_i$ of the four half-BPS operators~\cite{Eden:2012tu,Chicherin:2015edu}. They are given by a linear combination of conformal integrals with the coefficients depending 
on the choice of the $R-$charge polarization (or, 
equivalently, the $y-$variables) of the operators. Up to three-loop order, the emerging conformal integrals fall into two different classes -- `simple' ladder-type integrals, expressible in terms of classical 
polylogarithms, and `complicated' integrals, given by lengthy expressions involving Goncharov's polylogarithms. Examining the coefficients in front of the 
complicated integrals, one finds that, in application to \re{G4},  there exists a special choice of the polarizations of the half-BPS operators (to be specified shortly) for which these coefficients vanish 
simultaneously up to $K$ loops.  In this case, the weak coupling expansion of $G_4$ only involves the ladder integrals  up to order $g^{2K}$ and the complicated conformal integrals contribute starting 
from the order $O(g^{2K+2})$. This suggests that in the limit $K\to\infty$, the corresponding expressions for $G_4$ should simplify significantly to any order in the weak coupling expansion.

The above properties of the perturbative expansion for $G_4$ can be understood using the dual description of four-point correlation functions within the hexagonalization approach 
\cite{Basso:2015zoa,Fleury:2016ykk,Eden:2016xvg}. It takes full 
advantage of integrability of planar $\mathcal N=4$ SYM and allows one to obtain an equivalent expression for $G_4$, valid for any value of 't Hooft coupling, as a correlation function
of four hexagon operators $\vev{\mathcal H_1\mathcal H_2\mathcal H_3\mathcal H_4}$ in some integrable two-dimensional theory describing excitations propagating on a string world-sheet 
in the AdS/CFT. In the limit of infinitely heavy operators $K\to\infty$, for a special choice of the $R-$charge polarization,
this two-dimensional correlation function factorizes into the product of two-point functions of hexagon operators $\vev{\mathcal H_1 \mathcal H_2} \vev{\mathcal H_3 \mathcal H_4}$, see 
Ref.~\cite{Coronado:2018ypq}.  The latter have been dubbed the octagons, $\mathbb O_\ell=\vev{\mathcal H_1 \mathcal H_2}$. By inserting a complete set of intermediate states, the octagon 
can be expressed as a sum of the product of two hexagon form factors
\begin{align}\label{O-ff}
\mathbb O_\ell=\vev{\mathcal H_1 \mathcal H_2} = \sum_\psi \vev{\mathcal H_1 | \psi} \e^{-E_\psi \ell} \vev{\psi | \mathcal H_2}\,,
\end{align}
where $E_\psi$ is the (mirror) energy of the excitation $\psi$~\footnote{In the hexagonalization approach, these are elementary magnons and their bound states propagating in the mirror channel.} 
and the bridge length $\ell$ is given by the number of the scalar propagators stretched between the operators in $G_4$ (see Figure~\ref{fig:frames}). The properties of the magnon excitations as 
well as their hexagon form factors are known explicitly from integrability for any 't Hooft coupling \cite{Basso:2015zoa}.

For the four-point correlation functions described by Feynman diagrams shown in Figure~\ref{fig:frames}, the hexagonalization approach at large $K$ yields, schematically,
\begin{align}\label{s+a}
\mathcal G^{(s)} \sim [\mathbb O^{(s)}_{\ell=0}]^2 \,,\qqqquad
\mathcal G^{(a)} \sim \sum_{\ell=0}^{K/2} [\mathbb O^{(a)}_\ell]^2 \,.
\end{align}
Following Ref.~\cite{Coronado:2018ypq}, we shall refer to them as the \textit{simplest} and \textit{asymptotic} correlators, respectively. Similarly to the reduced correlation function in \re{G4}, the octagon 
depends on the 't Hooft coupling and two pairs of cross-ratios, $(z,\bar z)$ and $(\alpha,\bar\alpha)$. For the correlation functions \re{s+a}, corresponding values of $\alpha$ and $\bar\alpha$ are 
fixed by the choice of the $R-$charge polarization of the half-BPS operators~\footnote{In the notations of Ref.~\cite{Coronado:2018ypq}, these two relations correspond, respectively, to 
$\mathcal X^+=\mathcal X^-=-(1-z)(1-\bar z)$ and $\mathcal X^+=\mathcal X^-=1-(1-z)(1-\bar z)$, with $\mathcal X^+=-(z-\alpha)(\bar z-\alpha)/\alpha$ and $\mathcal X^-=-(z-\bar\alpha)(\bar z-\bar\alpha)/\bar\alpha$.}
\begin{align}\label{alpha-s}
& \alpha^{(s)}=\bar\alpha^{(s)}=1
\,,\qqqquad \alpha^{(a)}=\bar\alpha^{(a)}=\frac12 \left( z\bar z+\sqrt{(z\bar z)^2-4 z \bar z}\right)\,.
\end{align}
A somewhat unusual feature of the `asymptotic' correlation function is that the $R-$charge polarization of the half-BPS operators
involved depends on the space-time coordinates.

The relation \re{O-ff} provides a nonperturbative definition of the octagon in planar $\mathcal N=4$ that is valid for arbitrary 't Hooft coupling. It contains
an infinite sum  over excitations propagating on the world-sheet of the octagon across the bridge of length $\ell$. Direct evaluation of this sum for finite coupling is an extremely complicated task. 
However, it becomes feasible both at weak and strong coupling.

At weak coupling, for $g^2<1$, the contribution of an $n-$particle state to \re{O-ff} scales as $O(g^{2n(n+\ell)})$.  As a consequence, to any loop order,
$\mathbb O_\ell$ receives contributions only from a finite number of magnons. A calculation shows that 
$\mathbb O_\ell$ is given by a multi-linear combination of ladder integrals~\cite{Coronado:2018ypq,Coronado:2018cxj,Kostov:2019auq}. Its substitution into \re{s+a} yields a perturbative
series in $g^2$ that agrees for $K\to\infty$ with the results of Refs.~\cite{Eden:2012tu,Chicherin:2015edu} mentioned above.

At strong coupling, for $g^2\gg 1$, the sum in \re{O-ff} can be carried out using the clustering procedure developed in \cite{Jiang:2016ulr} in application to three-point correlation functions of non-protected
operators. The resulting expression for the octagon looks as $\mathbb O_\ell \sim \e^{-g A_0+O(g^0)}$ with some function $A_0$ depending on the cross-ratios.~\footnote{For large bridge 
length $\ell=O(g)$, the function $A_0$ also depends on $\bar\ell=\ell/(2g)$.} It possesses a typical semiclassical behavior anticipated within the AdS/CFT with $A_0$ having the meaning of the minimal area 
of a string that ends on four BMN-like geodesics \cite{Bargheer:2019kxb,Bargheer:2019exp}.

At finite coupling, the sum in \re{O-ff} can be cast as a determinant of a semi-infinite matrix~\cite{Kostov:2019stn,Kostov:2019auq}. In our previous paper, we used this result to show that the octagon 
at zero-length bridge $\mathbb O_{\ell=0}$ can be expressed as the Fredholm determinant of the (integrable) Bessel kernel. 
This representation proved to be very useful because it allowed us to apply powerful methods of integrable models 
to computing the octagon. In the next section, we review the findings of Refs.~\cite{Belitsky:2019fan,Belitsky:2020qrm} 
and extend that consideration to the octagons with arbitrary $\ell$. 
 
\section{Octagon as Fredholm determinant of Bessel operator}\label{sect:oct}

In this section, we recall the determinant representation of the octagon derived in Refs.~\cite{Kostov:2019stn,Kostov:2019auq} and recast it as the Fredholm determinant of a certain integral operator, schematically,
\begin{align}\label{O-Fred}
\mathbb O_\ell =  \vev{\mathcal H_1 \mathcal H_2} \sim \det(1 - \mathbf{H})\,.
\end{align}
This relation should not be surprising since two-point functions in two-dimensional integral models admit similar representation in terms of Fredholm determinants, see, e.g., Ref.~\cite{Korepin:1993kvr}. A notable 
example is the one-dimensional impenetrable Bose gas, in which case the integral operator $\mathbf H$ coincides with the Wiener-Hopf operator (see Eq.~\re{WH} below). We demonstrate below that for the 
octagon, the operator $\mathbf H$ in \re{O-Fred} coincides with the truncated Bessel operator. 

\subsection{Determinant representation}

As was shown in Refs.~\cite{Kostov:2019stn,Kostov:2019auq}, the sum over the intermediate states in \re{O-ff} yields the
representation  
\begin{align}\label{oct-gen}
\mathbb O_\ell(z,\bar z,\alpha, \bar\alpha) = \frac12 \sqrt{\det(1-\lambda_+ C K)} +  \frac12 \sqrt{\det(1-\lambda_- C K)}\,,
\end{align}
where $C_{mn} = \delta_{n+1,m}-\delta_{n,m+1}$ and $K_{mn}$ are semi-infinite matrices  (with $m,n=0,1,\dots$) and $\lambda_\pm= 2 \left[\cos\phi-\cosh(\varphi\pm i\theta)\right] $ are scalar factors. 

The matrix elements $K_{mn}$ are given by the integral involving the product of two Bessel functions
\begin{align}\label{K-mat}
 K_{mn} = {g\over 2i} \int_{|\xi|}^\infty dt  {\lr{i\sqrt{t+\xi\over t-\xi}}^{m-n}- \lr{i\sqrt{t+\xi\over t-\xi}}^{n-m}\over \cos\phi-\cosh t} J_{m+\ell}(2g\sqrt{t^2-\xi^2})J_{n+\ell}(2g\sqrt{t^2-\xi^2})\,.
\end{align}
The expressions for $\lambda_\pm$ and $K_{mn}$ together depend on the coupling constant $g$, bridge length $\ell$ and four variables $\xi,\phi,\varphi,\xi$. The latter are related to the cross-ratios \re{x-ratio} 
according to
\begin{align}\label{dict}
z=\e^{-\xi+i\phi}\,,\qquad \bar z= \e^{-\xi-i\phi}\,,\qquad \alpha = \e^{\varphi-\xi+i\theta}\,,\qquad \bar\alpha = \e^{\varphi-\xi-i\theta}\,.
\end{align}
The variables $\xi,\varphi,\theta$ take real values. The variable $\phi$ is real in Euclidean signature, whereas in the Lorentzian signature, it is convenient to switch to the variable $y$ instead
\begin{align}\label{y}
\phi=\pi + i y\,.
\end{align}
Expanding \re{oct-gen} in powers of the matrix $CK$, one finds that the terms involving $n$ such matrices describe the contribution to \re{O-ff} from $n$ magnons and their bound states.

As was demonstrated in Refs.~\cite{Belitsky:2019fan,Belitsky:2020qrm}, the matrix $H_\pm=\lambda_\pm CK$ has interesting properties.~\footnote{In these papers, the matrix $H$ was studied for zero-length
bridge $\ell=0$. The same analysis can be repeated verbatim for arbitrary $\ell$.} 
Namely, it can be brought to a block-diagonal form by an appropriate similarity transformation
\begin{align}\label{Omega}
H_\pm = \Omega^{-1} \left(\begin{array}{cc} k_\pm & 0 \\0 & k'_\pm \end{array}\right) \Omega \,,\qqqquad
k'_\pm =U^{-1}  k_\pm \, U \,,
\end{align}
where $k_\pm$ and $k'_\pm$ are semi-infinite matrices. The matrices $k_\pm$ are defined below in \re{k-pm}. The matrices $\Omega$ and $U$ can be found in Appendix~\ref{app:sim}, their  explicit form  is not important for our purposes.
Applying the relation \re{Omega}, we can simplify the expression for the octagon \re{oct-gen} as follows
\begin{align}\label{oct-gen1}
\mathbb O_\ell(z,\bar z,\alpha, \bar\alpha) = \frac12 \det(1-k_+) +  \frac12 \det(1-k_-)
\equiv \vev{\det(1-k)}\,,
\end{align}
where $\vev{\dots}$ denotes an average over $k=k_+$ and $k=k_-$.

The matrix elements $(k_\pm)_{mn}$ (with $m,n=0,1,2\dots$) are given by
\begin{align}\label{k-pm} 
 (k_\pm)_{nm} = (-1)^{n+m} (2n+\ell+1)\int_0^\infty {dx\over x} J_{2n+\ell+1}(\sqrt x) J_{2m+\ell+1}(\sqrt x)\chi_\pm(x) \,,
\end{align}
with the functions $\chi_\pm(x)$ being
\begin{align}\label{chi-pm}
 \chi_\pm(x) = {\cosh y + \cosh(\varphi\pm i\theta) \over \cosh y + \cosh\sqrt{x/(2g)^2 + \xi^2}}\,.
\end{align}
These functions decay exponentially fast at large $x$ and play the role of an ultraviolet cutoff in the integral \re{k-pm}. As we demonstrate below, the asymptotic behavior of the octagon at strong coupling is 
controlled by the behavior of $\chi_\pm(x)$ around the origin $x=0$.

Notice that the dependence of $k_\pm$ on the coupling constant and the kinematical variables $y,\xi,\varphi,\theta$ resides only in $\chi_\pm(x)$. 
For real variables, $\chi_+(x)$ and $\chi_-(x)$ are complex conjugate to each other. The same applies to the matrix elements $(k_+)_{nm}$ and $(k_-)_{nm}$ as well as to the Fredholm determinants of 
these matrices in \re{oct-gen1}. As a consequence, the octagon $\mathbb O_\ell(z,\bar z,\alpha, \bar\alpha)$ takes real values. 
 
For the `simplest' and `asymptotic' correlation functions defined in the previous section, the octagons on the right-hand side of \re{s+a} are given by a general expression \re{oct-gen1} with the variables 
$\alpha$ and $\bar\alpha$ being replaced by their values \re{alpha-s}. Using \re{dict}, we find that Eq.\ \re{alpha-s} translates to
\begin{align}
\cosh\varphi^{(s)}=\cosh\xi\,,\qqquad \cosh\varphi^{(a)} = \frac12\e^{-\xi}\,,\qqquad \theta^{(s)}=\theta^{(a)}=0\,.
\end{align}
Substituting these expressions into Eqs.~\re{oct-gen1} -- \re{chi-pm}, we obtain the following representation for the corresponding octagons $\mathbb O^{(s)}$ and $\mathbb O^{(a)}$
\begin{align}\label{oct-i}
\mathbb O^{(i)}_\ell(z,\bar z,\alpha, \bar\alpha) = \det(1-k^{(i)}) \,,
\end{align}
with $i=a,s$. Here, the semi-infinite matrices $k^{(s)}$ and $k^{(a)}$ are determined by Eq.\ \re{k-pm} with $\chi_\pm(x)$ replaced by the functions $\chi^{(s)}(x)$ and $\chi^{(a)}(x)$, respectively,
which read
\begin{align}\label{chi}\notag
\chi^{(s)}(x) = {\cosh y + \cosh \xi \over \cosh y + \cosh\sqrt{x/(2g)^2 + \xi^2}}\,,
\\
\chi^{(a)}(x) = {\cosh y + \frac12 \e^{-\xi} \over \cosh y + \cosh\sqrt{x/(2g)^2 + \xi^2}}\,.
\end{align}
We recall that, according to \re{dict} and \re{y}, the variables $y$ and $\xi$ are related to the cross-ratios 
of the space-time distances \re{x-ratio}   as
\begin{align}\label{y-xi}
y=-\frac12 \log(z/\bar z)\,,\qqqquad \xi = -\frac12\log(z\bar z) \,.
\end{align}
Notice that $\chi^{(s)}(0)=1$ and $\chi^{(a)}(0)= (\cosh y + \frac12 \e^{-\xi} )/ (\cosh y + \cosh \xi)<1$ for real $y$. As innocent as it might look, this property will play an important role in what follows. 

\subsection{Bessel operator}

In this subsection, we show that the determinants of the semi-infinite matrices entering Eqs.\ \re{oct-gen1} and \re{oct-i} can be expressed 
as the Fredholm determinants of the so-called Bessel operator.

Let us consider $\log \det(1-k) = -\sum_{n\ge 1} \tr (k^n)/n$, where a semi-infinite matrix $k$ takes the form \re{k-pm} with $\chi_\pm(x)$ replaced with some function $\chi(x)$. Applying \re{k-pm}, we can 
express  $\tr(k^n)$ as an $n-$fold integral,
\begin{align}\label{tr}
\tr(k^n) = \int_0^\infty dx_1 \dots \int_0^\infty dx_n \, K_\ell(x_1,x_2)\chi(x_1) \dots K_\ell(x_n,x_1)\chi(x_n)\,,
\end{align}
where $K_\ell(x,y)$ is given by 
\begin{align}\notag\label{K-ker}
K_\ell(x,y) 
&= {1\over \sqrt{xy}} \sum_{n=0}^\infty(2n+\ell+1)  J_{2n+\ell+1}(\sqrt x) J_{2n+\ell+1}(\sqrt y)
\\
&={\sqrt{x}\,J_{\ell+1}(\sqrt{x})J_{\ell}(\sqrt{y})-\sqrt{y}\,J_{\ell+1}(\sqrt{y})J_{\ell}(\sqrt{x})\over 2(x-y)}\,.
\end{align}
This function is known in the literature as the Bessel kernel, see, e.g., Refs.~\cite{Forrester:1993vtx,Tracy:1993xj}. 

Defining an integral operator with the kernel \re{K-ker}~\footnote{
The operator \re{K-op} is related to a self-adjoint operator with the kernel $K_\ell(x,y) (\chi(x) \chi(y))^{1/2}$ by a similarity transformation.} 
\begin{align}\label{K-op}
\mathbf{K}_\ell(\chi) f(x) =\int_0^\infty dy\, K_\ell(x,y) \chi(y) f(y)\,,
\end{align}
for a test function $f(x)$, the expression on the right-hand side of \re{tr} can be written as $\tr(\mathbf{K}_\ell^n)$. This allows us to express $\mathbb O_\ell(\chi)\equiv\det(1-k)$ 
as the Fredholm determinant of the operator \re{K-op} 
\begin{align}\label{TW}
\mathbb O_\ell(\chi)= \det\big(1- \mathbf K_\ell(\chi)\big)_{[0,\infty)} \,.
\end{align}
Here we inserted the subscript to indicate that the operator \re{K-op} acts on the positive half-axis. 

The determinant in \re{TW} depends on the choice of the cut-off function $\chi(x)$. For the octagons \re{oct-gen1} and \re{oct-i}, its form is fixed by the relations \re{chi-pm} and \re{chi}, e.g., 
\begin{align}
\mathbb O_\ell^{(a)} = \mathbb O_\ell(\chi^{(a)})\,,\qqqquad
\mathbb O_\ell^{(s)} = \mathbb O_\ell(\chi^{(s)})\,.
\end{align}
For the special choice $\chi(x) = \theta(s-x)$, the determinant \re{TW} previously appeared in the study of level spacing distributions in the Laguerre ensemble in random matrix theory 
\cite{Forrester:1993vtx,Tracy:1993xj}. In this case, the operator $\mathbf K_\ell$ acts on the interval $[0,s]$ and the Fredholm determinant \re{TW} gives the probability that no eigenvalues 
belong to it. As was shown in Ref.~\cite{Tracy:1993xj}, it can be found exactly in terms of a Painlev\'e transcendent. We argue in Section~\ref{sect:Toda}, that the same quantity governs 
the asymptotic behavior of the octagon at weak coupling in the limit when the four operators in \re{G4} are located at the vertices of a null rectangle.
 
We can obtain yet another representation for \re{TW} by using the following integral form of the kernel \re{K-ker} (see, e.g., Ref.~\cite{Tracy:1993xj})
\begin{align}
K_\ell(x,y) = \frac14 \int_0^1 dt \, J_\ell(\sqrt{xt})J_\ell(\sqrt{yt})\,.
\end{align}
Substituting this relation into \re{K-op} and examining $\tr(\mathbf K_\ell^n)$, we find that it can be re-expressed as 
\begin{align}\label{K=B}
\tr(\mathbf K_\ell^n)=\int_0^1 dt_1 \dots \int_0^1 dt_n \, B_\ell(t_1,t_2) \dots B_\ell(t_n,t_1) \equiv \tr(\mathbf B_\ell^n)\,,
\end{align}
where $\mathbf B_\ell$ is an integral operator acting on the interval $[0,1]$. It is defined as
\begin{align}\notag\label{B-ker}
& \mathbf B_\ell(\chi)\, f(t) = \int_0^1 dt' \,B_\ell(t,t') f(t')\,,
\\
& B_\ell(t,t') = \frac14\int_0^\infty dx\, J_\ell(\sqrt{tx})\chi(x) J_\ell(\sqrt{t' x})\,.
\end{align}
The operator $\mathbf B_\ell(\chi)$ is known in the literature as the truncated Bessel operator and the function $\chi(x)$ is referred to as its symbol, see, e.g., Ref.~\cite{BasorEhrhardt03}. 

Using the relation \re{K=B}, we thus arrive at another equivalent representation of the determinant \re{TW}
\begin{align}\label{equiv}
\mathbb O_\ell(\chi)
 = \det\big(1- \mathbf B_\ell(\chi)\big)_{[0,1]}\,.
\end{align}
This relation holds for an arbitrary cut-off function $\chi(x)$. Replacing the latter with \re{chi-pm} and \re{chi}, we get from \re{TW} and \re{equiv} two equivalent representations of the octagon as 
Fredholm determinants of the operators $\mathbf K_\ell$ and $\mathbf B_\ell$. 

Each of these representations has it own advantages. As we demonstrated in Ref.~\cite{Belitsky:2019fan,Belitsky:2020qrm}, 
in the special case of $\ell=0$, the representation \re{TW} can be used to derive a system of integro-differential equations for the octagon that 
is valid for arbitrary 't Hooft coupling. At weak coupling, these equations can be solved perturbatively to obtain an explicit expression for the octagon to any desired loop order.
At strong coupling, they allowed us to derive an expansion in powers of $1/g$ up to a constant $O(g^0)$ term. The latter arises as an integration constant and its determination requires an 
additional input. It comes from applying the strong Szeg\H{o} limit theorem to \re{equiv}. We show in Section~\ref{sect:str}, that the constant $O(g^0)$ term mentioned above is unambiguously 
given by the Szeg\H{o}-Akhiezer-Kac formula.

\subsection{Continuous bridge length}

By definition, the bridge length $\ell$ equals the number of the scalar propagators stretched between two half-BPS operators (see, e.g., Figure~\ref{fig:frames}) and, therefore, it takes nonnegative 
integer values. Applying Eq.\ \re{equiv}, we can extend the definition of the octagon to continuous real values of $\ell$. 

According to \re{K-ker}, \re{K-op} and \re{B-ker}, the $\ell-$dependence of the operators $\mathbf K_\ell(\chi)$ and $\mathbf B_\ell(\chi)$ is carried by the Bessel functions $J_\ell(x)$.  Although the latter are well-defined 
for arbitrary $\ell$'s, the requirement for the integral in \re{B-ker} to converge in the vicinity of $x\to 0$ imposes the condition $\ell >-1$. 

The values $\ell=\pm 1/2$ are of special interest because the product of the octagons $\mathbb O_{1/2}(\chi)$ and $\mathbb O_{-1/2}(\chi)$ is related to the Fredholm determinant of a truncated Wiener-Hopf 
operator $\mathbf W(\chi)$
\begin{align}\label{OO=W}
\mathbb O_{-1/2}(\chi) \, \mathbb O_{1/2}(\chi) = \det\big(1-\mathbf W(\chi)\big)_{[-1,1]}\,.
\end{align}
The operator $\mathbf W(\chi)$ acts on the interval $[-1,1]$ and is defined as 
\begin{align}\notag\label{WH}
& \mathbf W(\chi) f(t) = \int_{-1}^1 dt'\,W(t-t') f(t')\,,
\\
& W(t-t') = \int_{0}^\infty {dz\over\pi}\,\cos(z(t-t')) \chi(z^2)\,.
\end{align}
The derivation of the relation \re{OO=W} can be found in Appendix~\ref{app:WH}.

It is interesting to note that, for the cut-off function $\chi(x)$ given by the Fermi-Dirac distribution, the Fredholm determinant on the right-hand side of \re{OO=W} coincides with a temperature dependent two-point 
correlation function in the one-dimensional impenetrable Bose gas~\cite{Korepin:1993kvr}. In addition, for $\chi(x)=\theta(s-x)$, which corresponds to the zero temperature limit of the Fermi-Dirac distribution, the same 
determinant defines the probability that an arbitrary interval of length $s$ contains none of the eigenvalues in a random matrix Gaussian unitary ensemble \cite{Mehta}.
 
The determinant of the Wiener-Hopf operator \re{WH} has been intensively studied using both the method of differential equations~\cite{Korepin:1993kvr} and the strong Szeg\H{o} limit theorem~\cite{Bttcher2006AnalysisOT}. We 
shall use these findings together with \re{OO=W} to test our predictions for the octagon.
 
\section{Octagon at weak coupling}\label{sect:weak} 

As was alluded to in Section~\ref{sect:corr}, the perturbative expansion of the octagon at weak coupling involves only ladder integrals $q_k$ defined in Eq.~\re{ladder} below. The hexagonalization 
procedure allows us to organize this expansion in the number of particles $n=0,1,\dots$ propagating across the bridge in the dual representation of the octagon  \re{O-ff}. The $n-$particle state starts to contribute 
to \re{O-ff} only at order $O(g^{2n(n+\ell)})$ and the weak coupling expansion of the octagon takes the form \cite{Coronado:2018ypq,Coronado:2018cxj,Kostov:2019auq}
\begin{align}\notag\label{sum-I}
& \mathbb O_\ell = 1+ \sum_{n\ge 1}\mathcal I_{\ell,n}\,,
\\
& \mathcal I_{\ell,n} = \sum_{L=n(n+\ell)}^\infty g^{2L}  \sum_{L_1+\dots + L_n=L} d_{L_1\dots L_k} q_{L_1} \dots q_{L_k}\,,
\end{align}
where the functions $\mathcal I_{\ell,n}$ are given by multilinear combinations of the ladder integrals with
the expansion coefficients $d_{L_1\dots L_k}$ (with $L_i\ge 0$) depending on the bridge length $\ell$.  

\subsection{Expansion in ladder integrals}

Explicit expressions for the functions $\mathcal I_{\ell,n}$ can be found using the determinant representation of the octagon \re{equiv} in a straightforward fashion. Namely, expanding the determinant in powers of the 
Bessel operator, we can identify $\mathcal I_{\ell,n}$ as a contribution to $\mathbb O_\ell$ containing exactly $n$ copies of the operator $\mathbf B_\ell(\chi)$, e.g.,
\begin{align}\notag\label{IIs}
& \mathcal I_{\ell,1}= - \tr \mathbf B_\ell\,,\qquad 
\\[1.5mm]\notag
& 
\mathcal I_{\ell,2}= \frac12 \left[(\tr \mathbf B_\ell)^2-\tr(\mathbf B_\ell^2)\right]\,,\qquad 
\\[1.5mm]
& \mathcal I_{\ell,3}= -\frac16\left[(\tr \mathbf B_\ell)^3-3\tr \mathbf B_\ell\tr(\mathbf B_\ell^2)+2\tr(\mathbf B_\ell^3)\right]\,,\quad \dots
\end{align}
where $\tr(\mathbf B_\ell^n)$ (with $n=1,2,\dots $) is defined in \re{K=B}.
For small $g$, it is convenient to replace $x=(2gz)^2$ in \re{B-ker} and use an equivalent representation for the Bessel kernel
\begin{align}\label{B'}
 B_\ell(t,t') = 2g^2 \int_0^\infty dz\, z \,\widehat\chi(z) J_\ell(2g z \sqrt{t}) J_\ell(2gz \sqrt{t'}) \,,\qquad \widehat\chi(z) = \chi(4g^2z^2)\,.
\end{align}
Notice that, according to \re{chi-pm} and \re{chi}, the function $\widehat\chi(z) = \chi(4g^2z^2)$ is independent of the coupling constant. 

Expanding the product of the Bessel functions in \re{B'} in powers of $g^2$, we can write $B_\ell(t,t')$ as a double series in $\sqrt{t}$ and $\sqrt{t'}$ with its coefficients proportional to the following integral
\begin{align}\label{qk}
q_k (\chi)= 2k \int_0^\infty \, dz z^{2k-1} \widehat \chi(z)\,,
\end{align}
with $k \ge 1$. Its substitution into \re{IIs} yields the expressions for $\mathcal I_{\ell,n}$ of the form \re{sum-I}.
For instance, 
\begin{align}\notag\label{calI}
 \mathcal I_{\ell,1} &= 
(g^2)^{\ell+1}\sum_{k=0}^\infty  g^{2 k} (-1)^{k+1} \binom{2 (\ell+k)}{k}\frac{ 
   q_{\ell+k+1}}{\Gamma^2 (\ell+k+2)}\,,
\\\notag
 \mathcal I_{\ell,2}
 &= (g^2)^{2(\ell+2)} \bigg[ \frac{(\ell+2)^2 q_{\ell+1} q_{\ell+3}-(\ell+1) (\ell+3)
   q_{\ell+2}^2}{( \Gamma(\ell+3)\Gamma(\ell+4))^2} 
\\\notag
&   \qqqquad \qquad
   + g^2 \frac{2 (\ell+1) (\ell+4) q_{\ell+2} q_{\ell+3}-2 (\ell+2)
   (\ell+3) q_{\ell+1} q_{\ell+4}}{(\ell+2) (\Gamma(\ell+2)\Gamma (\ell+5))^2} + O(g^4) \bigg]\,,
\\\notag
 \mathcal I_{\ell,3}& =(g^2)^{3(\ell+3)} \bigg[ \frac{ (\ell+1) (\ell+4)^2 q_{\ell+5}
   q_{\ell+2}^2+(\ell+2)^2 (\ell+5) q_{\ell+1}
   q_{\ell+4}^2}{(\ell+3)^3(\ell+4)^4 (\ell+5)^2 \Gamma^6
   (\ell+3)}
 \\\notag
 &  \qqqquad \qquad - \frac{ 
   2 (\ell+1) (\ell+5) q_{\ell+2} q_{\ell+3} 
   q_{\ell+4}+(\ell+2) (\ell+4) q_{\ell+1} q_{\ell+3} 
   q_{\ell+5}}{(\ell+2)^5(\ell+3)^4(\ell+4)^3 (\ell+5)^2 \Gamma^6
   (\ell+2)}
 \\
 &  \qqqquad  \qquad
 + \frac{(\ell+1) (\ell+2)^2
    q_{\ell+3}^3}{(\ell+4)^2 (\ell+5) \Gamma^6
   (\ell+4)}  + O(g^2)\bigg]\,.
\end{align}
Expressions for higher order coefficients are too lengthy. To save space, we do not present them here.
 
Obviously, the integral in \re{qk} depends on the choice of the cut-off function $\widehat \chi(z) = \chi((2gz)^2)$ defined in
\re{chi-pm} and \re{chi}. Since the functions $\chi_\pm(x)$ and $\chi^{(i)}(x)$ differ by an overall $x-$independent factor, 
the corresponding integrals $q_k^\pm=q_k( \chi_\pm)$ and $q^{(i)}= q_k(\chi^{(i)})$ are proportional to each other
\begin{align}\label{q-rat}
& q_k^\pm={\cosh y + \cosh(\varphi\pm i\theta) \over \cosh y + \cosh \xi}q_k^{(s)}
\,,\qqqquad q_k^{(a)}={\cosh y + \frac12 \e^{-\xi} \over \cosh y + \cosh \xi}q_k^{(s)}\,.
\end{align}
For $\chi=\chi^{(s)}(x)$ 
given by \re{chi},  the function $q_k^{(s)}$ can be expressed in terms of classical polylogarithms
\begin{align}\notag\label{ladder}
q_k^{(s)} {}&= 2k \int_0^\infty \, dz \, z^{2k-1}{\cosh y + \cosh \xi \over \cosh y + \cosh\sqrt{z^2 + \xi^2}}
\\
{}&={(1-z)(1-\bar z)\over z-\bar z} \sum_{m=0}^k (-1)^m {(2k-m)! k!\over (k-m)! m!} \log^m(z\bar z) \left[
{\rm Li}_{2k-m}(z) - {\rm Li}_{2k-m}(\bar z)\right]\,,
\end{align}
where $y$ and $\xi$ are related to the cross-ratios $z$ and $\bar z$ by the relation \re{y-xi}. The expression in the second line of \re{ladder} is known as the ladder integral at $k$ 
loops~\cite{Usyukina:1993ch}.

Substituting the relations \re{calI} -- \re{ladder} into Eq.\ \re{sum-I}, we can obtain the weak coupling expansion for the octagon to any order in the coupling and for arbitrary values of the kinematical 
variables $y,\xi,\theta,\varphi$. 

\subsection{Octagon in null limit}\label{sect:null}

The octagon has interesting properties in the Lorentzian limit $x_{12}^2,x_{13}^2,x_{24}^2,x_{34}^2\to 0$ when the four operators in \re{G4} approach the vertices of a null rectangle. In terms of the 
cross-ratios \re{x-ratio} and kinematical variables \re{y-xi}, this limit corresponds
to $z\to 0$ and $\bar z\to\infty$ with $z\bar z$ kept fixed, or equivalently 
\begin{align}\label{null}
y\to\infty \,,\qqqquad \xi=\text{fixed}\,.
\end{align}
The two remaining variables $\varphi$ and $\theta$ remain finite in this limit. 

It follows from \re{q-rat}, that the functions $q_k^\pm$ and $q_k^{(i)}$ all coincide as $y\to\infty$. Together with \re{sum-I} this implies that, at weak coupling, the octagon $\mathbb O_\ell$ has a universal 
asymptotic behavior in the null limit, independent of the cross-ratios $\alpha$ and $\bar\alpha$, or, equivalently, the variables $\varphi$ and $\theta$.
A close examination of \re{ladder} shows that the function $q_k^{(s)}$ scales in the limit \re{null} as $q_k^{(s)}\sim y^{2k}$, thus producing large perturbative corrections to \re{sum-I} enhanced by powers of $y^2$.

For the octagon with zero-length bridge,  $\mathbb O_{\ell=0}$, such corrections exponentiate 
\begin{align}
\log \mathbb O_{\ell=0} = -{\Gamma(g)\over 2\pi^2} y^2 + {C(g)\over 8} + g^2 \xi^2\,,
\end{align}
where the functions $\Gamma(g)$ and $C(g)$ are known exactly \cite{Belitsky:2019fan}. This relation implies that to any order in 
$g^2$, perturbative corrections to $\log \mathbb O_{\ell=0}$ scale 
as $y^2$ in the null limit \re{null}. This property does not hold for $\ell\ge 1$, though. 

For the octagon with bridge length $\ell\ge 1$, weak coupling corrections to $\log \mathbb O_{\ell}$  scale in 
the null limit as $(g^2 y^2)^k$ with $k\ge \ell+1$. To resum them, we consider the following double scaling limit
\begin{align}\label{ds}
g \to 0 
\,,\qqquad
y\to\infty\,,\qqquad
s = 2g y =\text{fixed}\,.
\end{align}
Excluding the coupling constant in favor of $s$, we can re-expand $\log\mathbb O_{\ell}$ in powers of $1/y$ and determine accompanying $s-$dependent coefficient functions. Neglecting corrections 
subleading in $1/y$, we get
\begin{align}\label{s-exp}\notag
& \log  {\mathbb O}_0 = -{s^2\over 4}\,,
\\\notag
& \log  {\mathbb O}_1 = -\frac{s^4}{64}+\frac{s^6}{576}-\frac{11 s^8}{49152}+\frac{19
   s^{10}}{614400}+O(s^{12})\,,  
\\
& \log  {\mathbb O}_2 =-\frac{s^6}{2304}+\frac{s^8}{24576}-\frac{7
   s^{10}}{3686400}-\frac{s^{12}}{26542080}+O\left(s^{14}\right) \,,  \ \dots
\end{align}
For arbitrary $\ell$, the expansion of $\log {\mathbb O}_\ell$ starts at order $O(s^{2(\ell+1)})$ and runs in powers of $s^2$.
A question arises whether these series can be summed up to all orders in $s^2$.
 
\subsection{Toda lattice equations}\label{sect:Toda} 

Let us show that, in the double scaling limit \re{ds}, the octagon $\mathbb O_\ell$ satisfies nontrivial finite-difference relations known as the Toda lattice equations.

We start with the determinant representation \re{TW} and notice that the cut-off functions \re{chi-pm} and \re{chi} simplify  in the double scaling limit \re{ds} and approach
\begin{align}
\chi(x) \sim {1\over 1+\e^{(\sqrt x-s)/(2g)}}\,,
\end{align}
where we replaced the hyperbolic functions by their leading asymptotic behavior. Remarkably, this expression is the one of the Fermi-Dirac distribution with 
$T=2g$ and $\mu=s$ taking on the meaning of the temperature and chemical potential, respectively. The strict double scaling limit \re{ds} corresponds to the zero temperature 
when $\chi(x)$ reduces to the step function $\theta(s-\sqrt{x})$. 

Substituting $\chi(x) = \theta(s-\sqrt{x})$ into \re{K-op} and \re{TW}, we find that the octagon in the double scaling limit is given by the Fredholm determinant of an integral 
operator with the Bessel kernel \re{K-ker} acting on the finite interval $[0,s^2]$
\begin{align}\label{F}
\mathbb O_\ell = \det\big(1- \mathbf K_\ell\big)_{[0,s^2]}\,,
\end{align}
where the operator $\mathbf K_\ell$ is defined by Eq.\ \re{K-op} with $\chi(x) = \theta(s^2-x)$. As was already mentioned, the same Fredholm determinant defines the probability that no 
eigenvalues lie on the interval $[0,s^2]$ in the Laguerre unitary ensemble. For arbitrary $\ell$, it can be computed in terms of a Painlev\'e transcendent~\cite{Tracy:1993xj}. 

For nonnegative integer $\ell$, the Fredholm determinant in \re{F} can be expressed in terms of the modified Bessel functions~\cite{Forrester94,Forrester00}
\begin{align}\label{tau}
\mathbb O_\ell = \e^{-s^2/4} \tau_\ell(s)\,,\qqqquad 
\tau_\ell(s)= \det \Big[ I_{j-k}(s) \Big]_{j,k=1,\dots,\ell}\,.
\end{align}
This relation takes into account all perturbative corrections to the octagon  of the form $(g^2y^2)^k$ in the null limit.
For lowest values of $\ell$, we have
\begin{align}
\tau_0=1\,,\qqquad \tau_1= I_0(s)\,,\qqquad \tau_2=I_0^2(s) - I_1^2(s)\,,\qquad  \dots
\end{align}
The function $\tau_\ell(s)$ is closely related to the $\tau-$function in the Okamoto's theory of the Painlev\'e V equation~\cite{Okamoto,tau02,Forrester02}. As such, it satisfies a finite-difference relation, which
reads
\begin{align}\label{Toda}
s^2 { \tau_\ell\,  \tau_{\ell+2} \over  \tau_{\ell+1}^2 }
= (s\partial_s)^2 \log \tau_{\ell+1}\,.
\end{align}
It coincides with the Toda lattice equation.~\footnote{Changing variables, $s=\e^t$ and $\tau_{\ell+1}/\tau_{\ell}=\exp(q_\ell(t) -2 \ell t)$, one finds from \re{Toda} that $q_\ell(t)$ satisfies equations of motion of the Toda lattice, $\partial_t^2 q_\ell = \e^{q_{\ell+1}-q_\ell}- \e^{q_{\ell}-q_{\ell-1}}$.} 
We immediately verify that the expansion of \re{tau} at small $s$ agrees with \re{s-exp}. We also checked that the expressions \re{s-exp} satisfy the relation \re{Toda}.

At large $s$, the asymptotic behavior of the Fredholm determinant  \re{F} was found in Ref.~\cite{Tracy:1993xj}
\begin{align}\label{TW-as}
\mathbb O_\ell = \e^{-s^2/4+\ell s+c_\ell } s^{-\ell^2/2} \bigg[ 1+ {\ell\over 8} s^{-1} + {9 \ell^2 \over 128} s^{-2} + \lr{{3\ell\over 128} +{51\ell^3\over 1024}}s^{-3} +\dots\bigg] \,,
\end{align}
where the Dyson-Widom constant $c_\ell=\log(G(1+\ell)/(2\pi)^{\ell/2})$ is expressed in terms of the Barnes $G-$function. Its value was first conjectured in \cite{Tracy:1993xj} and later 
proved in \cite{EHRHARDT20103088}. It is convenient to rewrite \re{TW-as} as
\begin{align}\label{SAK-weak}
\log \mathbb O_\ell = -{s^2\over 4}+\ell s-\frac{\ell^2 }2\log s +  c_\ell +O(1/s)\,.
\end{align}
Notice that the leading term in \re{SAK-weak} is independent of the bridge length. 
 
The relations \re{tau} -- \re{SAK-weak} describe the asymptotic behavior of the octagon in the double scaling limit \re{ds}.
By definition, the coupling constant is small in this domain. In the following sections, we compute the octagon  at strong coupling and  
compare it with \re{SAK-weak} in Section~\ref{sect:null1}. 
 
\section{Octagon at strong coupling}\label{sect:str} 

In this section, we study the octagon $\mathbb O_\ell$ at strong coupling. We use the Fredholm determinant
representation \re{equiv} to derive the first few terms of the strong coupling expansion of $\log \mathbb O_\ell$. 

\subsection{Semiclassical expansion}

In the AdS/CFT description, the four-point correlation function \re{G4} is identified with a scattering amplitude of four closed strings dual to the four half-BPS operators defining \re{G4}. 
It was argued in Refs.~\cite{Bargheer:2019kxb,Bargheer:2019exp} that, in the limit of infinitely heavy operators, i.e., for $K\to\infty$, this amplitude can be computed using semiclassical 
expansion in string theory. To leading order in coupling, it is given by the area of a classical folded string that is attached to four BMN geodesics connecting the points $x_i$ 
on the AdS boundary and rotates on the sphere. 

The representation of the correlation functions as a product of the octagons  \re{s+a} corresponds to the
factorization of the closed string scattering amplitude into two copies of an off-shell open string partition functions. 
As a result, the strong coupling expansion of the octagon is expected to have a typical semiclassical form
\begin{align}\label{O-semi}
\mathbb O_\ell =\exp \left(-g A_0 + \frac12 A_1^2 \log g + B +O(1/g) \right).
\end{align}
Here $A_0$ is the minimal area of a single sheet of the folded string. The two terms involving functions $A_1$ and $B$ describe quadratic fluctuations of the string world-sheet. The 
remaining terms in the exponent describe higher order quantum fluctuations. 
A direct calculation of the coefficient functions $A_0,A_1,B, \dots$ is impractical by the existing methods in the AdS/CFT toolkit. We show below that they can be found by employing 
a powerful technique based on the strong Szeg\H{o} limit theorem.

The leading term $A_0$ was computed in Ref.~\cite{Bargheer:2019exp} using the clustering procedure~\cite{Jiang:2016ulr}. Its explicit expression is given in \re{1st} below.
In the previous work \cite{Belitsky:2020qrm}, we worked out a systematic $1/g$ expansion of the `simplest' octagon $\mathbb O^{(s)}_{\ell=0}$ by 
applying the method of differential equations \cite{Its:1990,Korepin:1993kvr}. This method allowed us to compute analytically all coefficient functions in \re{O-semi} except $B$. 
The latter appears  as an integration constant for the system of integro-differential equations for the octagon and it has to be determined independently by other means. Our goal is 
to find the constant term $B$  and to extend the findings of Ref.~\cite{Belitsky:2020qrm} to the octagon \re{O-semi} with arbitrary bridge length $\ell$. 
 
\subsection{Szeg\H{o}-Akhiezer-Kac formula}
 
To find the asymptotic behavior of the octagon at strong coupling, we use its representation \re{equiv} as the Fredholm determinant of the truncated Bessel operator \re{B-ker}. The dependence 
on the coupling constant enters \re{equiv} through the cut-off functions $\chi(x)$, see Eqs.\ \re{chi-pm} and \re{chi}. As was pointed out in Section~\ref{sect:Toda}, these functions resemble the 
Fermi-Dirac distribution with the coupling constant playing the role of the temperature. In this way, the strong coupling expansion of the octagon is analogous to the high-temperature expansion 
of correlation functions in two-dimensional integrable models~\cite{Korepin:1993kvr}.
 
The dependence of $\chi(x)$ on the coupling constant can be eliminated by substituting $x=(2gz)^2$ and introducing a notation for $\widehat \chi(z) = \chi(4g^2 z^2)$.
According to \re{chi-pm} and \re{chi}, the function $\widehat \chi(z)$ does not depend on $g$ and is given by
\begin{align}\notag\label{hatchi}
& \widehat \chi_\pm(z) = {\cosh y + \cosh(\varphi\pm i\theta) \over \cosh y + \cosh\sqrt{z^2 + \xi^2}}\,,
\\\notag
&\widehat \chi^{(s)}(z) = {\cosh y + \cosh \xi \over \cosh y + \cosh\sqrt{z^2 + \xi^2}}\,,
\\
&\widehat \chi^{(a)}(z) = {\cosh y + \frac12 \e^{-\xi} \over \cosh y + \cosh\sqrt{z^2 + \xi^2}}\,.
\end{align}
Changing the integration variables in \re{B-ker} to $t=\tau/(2g)^2$ and $t'=\tau'/(2g)^2$, we can rewrite \re{equiv} as
\begin{align}\label{trB}
\mathbb O_\ell = \det\big(1-{\mathbf B}_\ell(\widehat{\chi})\big)_{[0,2g]}\,,
\end{align}
where the truncated Bessel operator ${\mathbf B}_\ell(\widehat{\chi})$ acts now on the interval $[0,2g]$ and is defined by
\begin{align}\notag\label{B-hat}
& {\mathbf B}_\ell(\widehat{\chi})\, f(\tau) = \int_0^{2g} d\tau' \,\widehat{B}_\ell(\tau,\tau') f(\tau')\,,
\\[2mm]
& \widehat{B}_\ell(\tau,\tau') = (\tau\tau')^{1/2} \int_0^\infty dz \, z J_\ell(\tau z)\widehat\chi(z) J_\ell(\tau' z)\,.
\end{align}
In distinction to \re{B-ker}, its kernel does not depend on $g$.

The asymptotics of the determinant \re{trB} at large $g$ was studied in Ref.~\cite{BasorEhrhardt03}. It was shown there that for real $\ell>-1$ and sufficiently smooth function $\widehat\chi(z)$,
such that $\widehat\chi(z)\neq 1$ for $z\ge 0$,
\footnote{In what follows, we refer to functions satisfying these conditions as `regular' symbols.} it is given by 
\begin{align}\label{SAK-B}
\det\big(1-{\mathbf B}_\ell(\widehat{\chi})\big)_{[0,2g]}
= \exp \bigg( 2g \widetilde \psi(0) - {\ell\over 2} \psi(0) + \frac12 \int_0^\infty dk\, k \big(\widetilde\psi(k)\big)^2 + O(1/g)\bigg)\,,
\end{align}
where a notation was introduced for
\begin{align}\label{psi}
\psi(z) = \log(1-\widehat\chi(z))\,,\qqqquad
\widetilde \psi(k) = \int_0^\infty {dz\over\pi} \cos(kz) \log(1-\widehat\chi(z))\,.
\end{align}
The relation \re{SAK-B} is similar to the well-known Szeg\H{o}-Akhiezer-Kac formula for determinants of truncated Wiener-Hopf operators~\cite{Kac:1964,Akhiezer:1964}. The last term in the exponent 
of \re{SAK-B} stands for corrections vanishing as $g\to\infty$. We discuss them in Section~\ref{sect:exp} below.
 
We observe a remarkably similarity of expressions in the right-hand side of Eqs.\ \re{O-semi} and \re{SAK-B}. Matching the two relations, we deduce that
\begin{align}\notag\label{1st}
& A_0= -2\int_0^\infty {dz\over\pi} \log(1-\widehat\chi(z))\,,
\\\notag
& A_1=0\,,
\\[1mm]
& B=- {\ell\over 2}\log(1-\widehat\chi(0)) + \frac12 \int_0^\infty dk\, k \big(\widetilde\psi(k)\big)^2\,,
\end{align}
where $\widetilde\psi(k)$ is defined in \re{psi}. The formula for $A_0$ is a generalization of the first Szeg\H{o} theorem~\cite{Szego:1915} to the truncated Bessel operator.
It coincides with the analogous expressions obtained in Refs.~\cite{Bargheer:2019exp,Belitsky:2020qrm} using different techniques.

We would like to stress that the relations \re{1st} hold for arbitrary real $\ell>-1$ and the function $\widehat\chi(z)$ verifying the conditions mentioned above (i.e., a regular symbol). 
In our previous work \cite{Belitsky:2020qrm}, we found that $A_1=1$ in the special case of $\ell=0$ and $\widehat\chi = \widehat\chi^{(s)}(z)$ (see Eq.~\re{hatchi}). This result 
obviously contradicts the second relation in \re{1st}. 

To understand the reason for this, we note that, according to its definition \re{hatchi}, the function $\widehat\chi^{(s)}(z)$ satisfies the relation $\widehat\chi^{(s)}(0)=1$ and, therefore, it does 
not fulfill the conditions for the validity of the relations \re{SAK-B} and \re{1st}.  Thus,  $\widehat\chi^{(s)}(z)$ defines a singular symbol. Indeed, a naive substitution of $\widehat\chi^{(s)}(z)$ 
into the last relation in \re{1st} yields a logarithmic divergence. We show in the next subsection, that careful analysis of the determinant with the singular symbol 
leads to a finite expression, in which this divergence gets replaced by a $\log g$ contribution, thus generating a nonzero result for $A_1$, in agreement with the finding of Ref.\
\cite{Belitsky:2020qrm}. 

In counter-distinction to $\widehat\chi^{(s)}(z)$, the function $\widehat\chi^{(a)}(z)$ satisfies $\widehat\chi^{(a)}(z)<1$ and, therefore, it defines a regular symbol. In this case, the 
relations \re{1st} describe the asymptotic behavior of the asymptotic octagon $\mathbb O_\ell^{(a)}=\det(1-{\mathbf B}_\ell(\widehat{\chi}^{(a)}))$ at large $g$. For the functions 
$\widehat\chi^{\pm}(z)$, the requirement $ \widehat\chi^{\pm}(z)\neq 1$ leads to nontrivial restrictions for the kinematic variables: $\sin\theta \sinh\varphi\neq 0$ or 
$\cosh\xi>  \cos\theta\cosh\varphi$, otherwise. The relations \re{1st} are applicable to the octagon $\mathbb O_\ell^{\pm}$ as long as these conditions are satisfied.
For the sake of simplicity we assume that this is the case.
 
\subsection{Fisher-Hartwig singularities}\label{sect:FH} 

We demonstrated in the previous subsection, that the Szeg\H{o}-Akhiezer-Kac formula \re{SAK-B} has to be modified for 
$\widehat\chi=\widehat\chi^{(s)}(z)$. At small $z$, we find from \re{hatchi}
\begin{align}\label{bs}
1-\widehat\chi^{(s)}(z)={z^2\sinh\xi\over 2\xi(\cosh\xi+\cosh y)} + O(z^4)\,.
\end{align}
Substituting this expression into \re{psi}, we deduce that $\widetilde \psi(k)\sim -1/k $ at large $k$. As a consequence, the second  term in the expression for $B$ in \re{1st} also diverges logarithmically. 
This divergence is a manifestation of a Fisher-Hartwig singularity~\cite{Fisher68} (see also \cite{Bttcher2006AnalysisOT}, for a review). 

An expression for the symbol with the Fisher-Hartwig singularity $\omega_\beta$ looks as~\footnote{A general expression 
for the symbol with the Fisher-Hartwig singularities also contains factors of form $[(z-i)/(z+i)]^\gamma$. They do not appear in our analysis.}
\begin{align}\label{FH}
1-\widehat\chi(z)= b(z) \omega_\beta(z) \,, \qqqquad  \omega_\beta(z)=\lr{z^2\over z^2+1}^\beta\,,
\end{align}
where $b(z)$ is a regular symbol and $\beta$ is an arbitrary parameter. Comparing this relation with \re{bs} and \re{hatchi}, we 
find that $\beta$ takes the following values for different functions
\begin{align}\label{betas}
\beta^{(s)}=1\,,\qqqquad \beta^{(a)}= \beta^\pm =0\,.
\end{align}
At present, the large $g$ asymptotics of the determinant of the Bessel operator \re{trB} with the singular symbol \re{FH} is known only for specific values of the index of the Bessel function $\ell=\pm 1/2$, see Ref.~\cite{BasorEhrhardt05}. 
In this case, the Bessel operator reduces to a sum or difference of the Wiener-Hopf and Hankel operators.
 
 For our purposes, we need an expression for the determinant of the Bessel operator for arbitrary $\ell$ with a singular symbol of the form \re{FH}. We argue in Appendix~\ref{app:HF}, that it has the following 
 conjectural form 
\begin{align}\notag\label{SAK}
\det\big(1-\mathbf B_\ell(\widehat\chi)\big)_{[0,2g]} {}& = \exp\bigg[ 2g \widetilde \psi(0) + \frac12 \int_0^\infty dk\, \left[ k(\widetilde \psi(k))^2-\beta^2 {1-\e^{-k}\over k}\right]
  -\frac{\ell}2\log b(0)
\\
{}&+
\lr{\beta\ell + \frac12\beta^2}\log g
+ \frac{\beta}2   \log(2\pi)  + \log {G(1+\ell)\over G(1+\ell+\beta)} + O(1/g)\bigg],
\end{align}
where the functions $\widetilde \psi(k)$ and $b(z)$ are defined in \re{psi} and \re{FH}, respectively, and $G$ is the Barnes function. It is easy to check using \re{psi} and \re{FH} that $\widetilde\psi(k) \sim -\beta/k$ 
at large $k$ and, therefore, the integral in the first line of \re{SAK} is well-defined.

The relation \re{SAK} generalizes the Szeg\H{o}-Akhiezer-Kac formula \re{SAK-B} to symbols with the Fisher-Hartwig singularity. The two relations coincide for $\beta=0$. To find the asymptotic behavior of the 
octagon $\mathbb O_\ell^{(s)}$, we substitute $\beta=1$ into \re{SAK} and match it to \re{O-semi}. In this way, we arrive at
\begin{align}\notag\label{2nd}
& A_0^{(s)} = -2\int_0^\infty {dz\over\pi} \log\left(1-\widehat\chi^{(s)} (z)\right)\,,
\\\notag
& A_1^{(s)} = (2\ell+1)^{1/2}\,,
\\[1mm]
& B^{(s)} = \frac12 \int_0^\infty dk\, \left[ k(\widetilde \psi(k))^2-{1-\e^{-k}\over k}\right] -\frac{\ell}2\log b(0)
 - \log {\Gamma(1+\ell)\over \sqrt{2\pi}} \,,
\end{align}
where $\widetilde\psi(k)$ is given by \re{psi} with $\widehat\chi=\widehat\chi^{(s)}(z)$ and
\begin{align}
b(0) =-\frac12\widehat\chi''(0) = {\sinh\xi \over 2\xi(\cosh \xi+\cosh y)}\,.
\end{align}
The following comments are in order.

Comparing \re{1st} and \re{2nd}, we observe that the leading term $A_0$ takes the same form for all three cut-off functions in \re{hatchi}. The subleading term $A_1$ in \re{2nd} is generated by 
the Fisher-Hartwig singularity. It is sensitive to the asymptotic behavior of the function $\widehat\chi(z)$ around the origin
and does not depend on the kinematical variables. We verify, that $A_1=1$ for $\ell=0$ is in agreement with the finding of Ref.~\cite{Belitsky:2020qrm}.
The last relation in \re{2nd} yields a prediction for the function $B$. We show in Appendix~\ref{app:B} that for $\ell=0$, it agrees with numerical results for this function obtained in Ref.~\cite{Belitsky:2020qrm}.

\section{Strong coupling expansion}\label{sect:exp}

In the previous section, we applied the strong Szeg\H{o} limit theorem to determine the first three terms in the strong coupling 
expansion of the octagon \re{O-semi} for three different cut-off functions (symbols) \re{hatchi}. For regular symbols $\widehat\chi_\pm$ and $\widehat\chi^{(a)}$, they are given by \re{1st} and, for the 
singular symbol $\widehat\chi^{(s)}$, by \re{2nd}. 

In this section, we apply the powerful method of differential equations \cite{Its:1990,Korepin:1993kvr} to compute subleading corrections to \re{O-semi} suppressed by powers of $1/g$
\begin{align}\label{O-str}
\log \mathbb O_\ell = -g A_0 + \frac12 A_1^2 \log g + B + \sum_{k\ge 2} {A_k\over 2k(k-1)} g^{1-k}\,,
\end{align}
where the rational factors are inserted for convenience. The expansion coefficients in \re{O-str} depend on 
the symbol \re{hatchi}. Applying \re{FH} and \re{SAK}, we can write 
expressions for the coefficients $A_0, A_1,B$ in a unified form by introducing the dependence on $\beta$
\begin{align}\notag\label{uni}
& A_0 = -2\int_0^\infty {dz\over\pi} \log\left(1-\widehat\chi(z)\right)\,,
\\\notag
& A_1^2 = 2\beta\ell + \beta^2\,,
\\[1mm]
& B =\frac12 \int_0^\infty dk\, \left[ k(\widetilde \psi(k))^2-\beta^2 {1-\e^{-k}\over k}\right]
+ \frac{\beta}2   \log(2\pi)  + \log {G(1+\ell)\over G(1+\ell+\beta)}  -\frac{\ell}2\log b(0) \,,
\end{align}
where $\widetilde \psi(k)$ is defined in \re{psi} and the variables $\beta$ and $b(0)$ control  the behavior of the symbol in the vicinity of the origin,  $\widehat\chi(z)\sim 1-b(0) z^{2\beta}$.
The relations \re{1st} and \re{2nd} correspond to $\beta=0$ and $\beta=1$, respectively.  

Notice that the leading term of the strong coupling expansion \re{O-str} does not depend on the bridge length $\ell$.
This dependence first appears in $O(\log g)$ term which is linear in $\ell$. The expansion in \re{O-str} is well-defined 
provided that $\ell$ stays finite as $g\to\infty$. We show below, that for $\ell=O(g)$ all terms in \re{O-str} scale as $O(g)$ 
and the series has to be resummed. We solve this problem in Section~\ref{sect:large}.
 
\subsection{Method of differential equations} 

In our previous work \cite{Belitsky:2019fan,Belitsky:2020qrm}, we applied the method of differential equations~\cite{Its:1990,Korepin:1993kvr} to compute the expansion coefficients in \re{O-str} for the `simplest' octagon 
$\mathbb O_\ell^{(s)}$ at zero length bridge $\ell=0$.
The analysis in this section goes along the same lines and we refer to the above mentioned papers for details.  
 
The method of differential equations allows us to obtain a system of exact equations for the so-called potential $u$, defined as a logarithmic derivative of the octagon,
\begin{align}\label{u}
u=-2 g \partial_g \log \mathbb O_\ell \,.
\end{align}
Substituting \re{O-str} into this relation, we obtain a generic expression for the potential at strong coupling
\begin{align}\label{u-strong}
u= 2g A_0 - A_1^2 + \sum_{k\ge 2} {A_k\over k} g^{1-k}\,.
\end{align}
Notice that the function $B$ does not contribute to $u$.  

Using the representation of the octagon as the Fredholm determinant of the Bessel operator \re{TW}, we can show that $u$ satisfies the following exact relation
\begin{align}\label{du}
 \partial_g u = -8g \int_0^\infty dz\, z^2 \, q^2(z) \partial_z \widehat \chi(z)\,,
\end{align}
where the function $q(z)$ is a solution to the differential equation
\begin{align}\label{q}
\lr{g\partial_g }^2 q (z) + \lr{4g^2 z^2-\ell^2 - g\partial_g u + u} q (z) = 0
\, ,
\end{align} 
subject to the boundary condition $q(z) = J_\ell(2gz) + O(g^{\ell+2})$ at weak coupling. 
Compared to the case $\ell = 0$  addressed in  \cite{Belitsky:2019fan,Belitsky:2020qrm}, the only modification is
the additional term $(-\ell^2)$ inside the brackets in \re{q}.

Being combined together, the relations \re{du} and \re{q} define a coupled system of equations for $u$ that is valid for any value of the coupling constant. At weak coupling, it is straightforward 
to solve \re{du} and \re{q} perturbatively and expand $u$ in powers of $g^2$. Together with \re{u}, this leads to a weak coupling expansion of the octagon $\mathbb O_\ell$
that coincides with \re{sum-I}.

\subsection{Quantization conditions}

At strong coupling, we replace the potential in \re{q} with its general expansion \re{u-strong} and take into account \re{uni} 
to get
\begin{align}\label{pde-start1}
\left[ \lr{g\partial_g }^2 + 4(gz)^2-(\ell+\beta)^2+{A_2 \over g} +{A_3 \over g^2}+\dots \right]q(z)=0\,.
\end{align}
To leading order, neglecting terms suppressed by powers of $1/g$ inside the brackets and requiring $q(z)$ to be regular for $g z\to 0$, we find that $q(z)$ is proportional to the Bessel function 
$J_{\ell+\beta}(2gz)$. Then, the coefficient in front of $J_{\ell+\beta}(2gz)$ is fixed by evaluating the integral in the right-hand side of \re{du} at large $g$ and matching it to the expected result 
\re{u-strong},  which reads $\partial_g u=2 A_0+O(1/g)$ with $A_0$ given by \re{uni}. 

This leads to
\begin{align}\label{q-f}
q(z) = {f(z,g)\over [1-\widehat\chi(z)]^{1/2}}\,,\qqqquad
f(z,g) = J_{\ell+\beta}(2gz) + O(1/g)\,.
\end{align}
At large $g z$, the Bessel function in the last relation behaves as $ (a_0\sin(2 gz ) + b_0\cos(2gz))/\sqrt{2\pi gz }$ with some constant $a_0$ and $b_0$.
To find subleading corrections to $q(z)$, we look for the function $f(z,g)$ in a similar form
\begin{align}\label{f-trig}
f(z,g) =  
{1\over \sqrt{2\pi gz }}
\Big[a(z,g) \sin(2 gz ) + b(z,g) \cos(2gz) \Big]
\, ,
\end{align}
where $a_0$ and $b_0$ are replaced by an infinite series in $1/g$ with $z-$dependent coefficients,
\begin{align}\label{ab}
a(z,g)= \sum_{n=0}^\infty  {a_{n} (z)\over (gz)^n} \,,\qqqquad b(z,g)= \sum_{n=0}^\infty  {b_{n} (z)\over (gz)^n} 
\, .
\end{align}   
The symmetry of the differential equation \re{pde-start1} under the reflection $z\to -z$ leads to $a (-z,g) = - b(z,g)$.
The coefficient functions $a_n(z)$ and $b_n(z)$ can be found by substituting Eqs.~\re{q-f}--\re{ab} into \re{pde-start1} and 
equating to zero the coefficients in front of powers of $1/g$. In this manner, we obtain 
\begin{align}\notag\label{a2}
& \frac12\left[a^2(z,g) + b^2(z,g)\right]  =
1+\frac{4\ell_\beta^2-1}{32 z^2 g^2 }-\frac{A_2}{8 z^2 g^3 }
\\[2mm] 
& \qquad  + \lr{\frac{3(4\ell_\beta^2-1)(4\ell_\beta^2-9)}{2048 z^4}-\frac{A_3}{8 z^2}}{1\over g^4}
 - \lr{\frac{3(4\ell_\beta^2-9 )A_2}{256 z^4}+\frac{A_4}{8
   z^2}}{1\over g^5}+O\left(1/g^6\right),
\end{align}
where the notation was introduced for $\ell_\beta=\ell+\beta$.
 
Substituting \re{q-f} and \re{f-trig} into \re{du}, we find that $\partial_gu$ is given by an integral involving rapidly 
oscillating trigonometric functions. Up to  corrections that are exponentially small at large $g$, we can 
replace these functions by their mean values to get from \re{du}
\begin{align}\label{mat}
\partial_g u = {2\over\pi} \int_0^\infty dz \, z \partial_z \log(1-\widehat\chi(z)) \left[a^2(z,g) + b^2(z,g)\right].
\end{align}
According to \re{u-strong}, the expression in the left-hand side is given by a series in $1/g$ with expansion 
coefficients proportional to $A_k$ (with $k=0,1,\dots$). Taking into account \re{a2}, we find that the right-hand side of
\re{mat} has a similar form. Matching the coefficients in front of powers of $1/g$ on both sides of \re{mat}, we can
determine the coefficients $A_k$ for any index $k$. 

\subsection{Expansion coefficients}

A first few coefficients are given by~\footnote{We
present expressions for the coefficients $A_k$ (with $2 \le k \le 30$) in an ancillary file
accompanying our paper. } 
\begin{align}\notag\label{As-beta}
A_2 {}&=-\frac{1}{4}  (4\ell_\beta^2 -1) I_1\,,
\\[1mm]\notag
A_3 {}&=
-\frac{3}{16}  (4\ell_\beta^2 -1) I_1^2\,,
\\[1mm]\notag
A_4 {}&=  
  -\frac{1}{128}  (4\ell_\beta^2 -1) \left(\left(4 \ell_\beta^2-9\right)I_2 +16
  I_1^3\right),
\\[1mm]\notag
A_5 {}&=  
  -\frac{5}{256} (4\ell_\beta^2 -1)\left(\left(4 \ell_\beta^2-9\right) I_1I_2  +4I_1^4\right), 
\\[1mm]\notag
A_6 {}& = - \frac{3}{8192} (4\ell_\beta^2 -1)\left(80 (4 \ell_\beta^2-9) I_2 I_1^2+(4
   \ell_\beta^2-9)(4 \ell_\beta^2-25)  I_3+128 I_1^5 \right), 
\\[1mm]\notag
A_7 {}& = -\frac{7}{16384}(4\ell_\beta^2 -1)\big(80 (4 \ell_\beta^2-9)I_2I_1^3+3 (4
   \ell_\beta^2-9)(4 \ell_\beta^2-25) I_3I_1
\\[1mm]\notag{}& \phantom{=}    
   + (4
   \ell_\beta^2-9)(4 \ell_\beta^2-21) I_2^2+64I_1^6\big),
\\[1mm]\notag
A_8{}&= -\frac1{262144}(4\ell_\beta^2 -1)\big(8960 (4 \ell_\beta^2-9)I_2I_1^4+672 (4 \ell_\beta^2-9) (4 \ell_\beta^2-25)I_3
  I_1^2
\\[1mm] {}& \phantom{=}    
+448 (4 \ell_\beta^2-9) (4 \ell_\beta^2-21)I_2^2 I_1+5  (4
   \ell_\beta^2-9) (4 \ell_\beta^2-25) (4 \ell_\beta^2-49)I_4+4096I_1^7\big), 
\end{align}
where, as before, $\ell_\beta=\ell+\beta$ and a notation was introduced for the so-called `profile' function,
\begin{align}\label{In}
I_n(\widehat\chi)= \int_0^\infty  {dz\over\pi}\, z^{1-2n} \partial_z \log(1-\widehat \chi(z) )\,.
\end{align}
For $\ell=0$ and $\beta=1$, or equivalently $\ell_\beta=1$, the relations \re{As-beta} coincide with analogous expressions in  Ref.~\cite{Belitsky:2020qrm},
see Eq.\ (5.18) there. 

According to \re{FH}, $\partial_z \log(1-\widehat \chi(z) )\sim 2\beta/z$ for $z\to 0$ and the integral in \re{In} is divergent for 
$n\ge 1/2$. As explained in Ref.~\cite{Belitsky:2020qrm}, the integral \re{In} has to be defined by analytically continuing $I_n$ from 
complex $n$ with $\Re{\rm e}\, n<1/2$ to positive integer $n$. In this manner, we obtain an equivalent representation for the profile function
\begin{align}\label{Ia} 
I_n(\widehat\chi) &= {1\over (2n-1)!!} \int_0^\infty {dz\over\pi} (z^{-1} \partial_z)^n z\partial_z \log(1-\widehat\chi(z))\,,
\end{align}
which is well-defined for arbitrary nonnegative $n$.
For $n=0$, the function \re{Ia} gives the leading term of the strong coupling expansion \re{uni}, $A_0=2I_0$.
The dependence on the kinematical variables enters $I_n$ through the cut-off function $\widehat\chi(z)$. Using \re{FH}, we
get
\begin{align}\label{Ibeta}
I_n(\widehat\chi) &=(-1)^n \beta  + I_n(1-b)\,,
\end{align}
where $ I_n(1-b)$ is given by \re{Ia} with $\widehat\chi(z)$ replaced by $1-b(z)$.

Being combined together, the relations \re{uni} and \re{As-beta} determine the strong coupling expansion of the octagon
\re{O-str} to arbitrary order in $1/g$. It has the following interesting properties. The coefficient $A_1$ accompanying $\log g$ term in \re{O-str} is independent of the 
kinematical variables, whereas the coefficients in front of powers of $1/g$ depend on the cut-off function $\widehat\chi(z)$. These coefficients grow factorially for generic 
values of the kinematical variables.  
Applying the standard resummation 
technique, we can use the strong coupling expansion to compute the octagon for finite coupling~\cite{Belitsky:2020qrm}. 

The expansion coefficients $A_{2k}$ and $A_{2k+1}$ (with $k\ge 1$) are even polynomial in $\ell_\beta$ of degree $2k$. They vanish for 
$\ell_\beta=\pm 1/2$ and simplify significantly for half-integer $\ell_\beta$. We show in the next section that in this case, the series \re{O-str} can be resummed to all orders in $1/g$. 
 
\section{Properties of strong coupling expansion}\label{sect:prop}

In this section, we use Eqs.\ \re{O-str}, \re{uni} and \re{As-beta} to demonstrate that certain type of corrections to the octagon
can be summed up to all orders in $1/g$, thus improving convergence properties of the strong coupling expansion.

\subsection{Resummation}

Following Ref.\ \cite{Belitsky:2020qrm}, we split the expression for the octagon \re{O-str} as
\begin{align} \label{O-split}
& \log \mathbb O_\ell  = -2 g I_0-\frac18(4\ell^2-1)\log g + B + \log \mathbb O_{\ell,q} \,,
\end{align}
where $\log \mathbb O_{\ell,q}$ is given by
\begin{align}\label{Oq}
 \log \mathbb O_{\ell,q} =\frac{1}{8} (4\ell_\beta^2-1) \log g  +\sum_{k\ge 2} {A_k\over 2k(k-1)} g^{1-k} \,.
\end{align}
The relation \re{O-split} holds up to corrections which are exponentially small at large $g$.

Notice that the dependence on the index $\ell$ of the Bessel function and the strength $\beta$ of the Fisher-Hartwig singularity enters \re{Oq} through their sum $\ell_\beta=\ell+\beta$. 
This is not the case for the coefficient function $B$
defined in \re{uni}.

According to \re{As-beta}, the $1/g-$series for $\log \mathbb O_{\ell,q}$ contains terms involving powers of $I_1$. All such terms can be eliminated by changing the 't Hooft coupling to
\begin{align}
g'=g-I_1/2\,.
\end{align}
The expansion of $\log \mathbb O_{\ell,q}$ in the shifted parameter $g'$ becomes
\begin{align}\notag\label{Oq-ser}
\log \mathbb O_{\ell,q}  &=\frac{1}{8} (4 \ell^2_\beta-1) \log ({g'})
-\frac{(4 \ell_\beta^2-1)(4 \ell_\beta^2-9)}{3072 {g'}^3}I_2
 \\\notag & 
 -\frac{ (4 \ell_\beta^2-1)(4 \ell_\beta^2-9)(4 \ell_\beta^2-25)}{163840 {g'}^5}I_3
  \\    {}& 
  -\frac{(4 \ell_\beta^2-1)
   (4 \ell_\beta^2-9) (4 \ell_\beta^2-21)}{196608
   {g'}^6}I_2^2+O(1/{g'}^7) 
   \, .
\end{align}
As a check, we verify that for $\ell_\beta=1$, this expression coincides with Eq.~(6.7) in \cite{Belitsky:2020qrm}.

It is remarkable that for half-integer $\ell_\beta$, the series \re{Oq-ser} can be resummed, e.g.,
\begin{align}\notag\label{exa}
\log \mathbb O_{\ell,q} \big|_{\ell_\beta=\pm \frac12} &= 0 \,,
\\[1.5mm]\notag
\log \mathbb O_{\ell,q} \big|_{\ell_\beta=\pm \frac32} &= \log(g')\,,
\\[1.5mm]\notag
\log \mathbb O_{\ell,q} \big|_{\ell_\beta=\pm \frac52} &= 3 \log(g') + \log\left(1-I_2/(2g')^3\right)\,,
\\[1.5mm]
\log \mathbb O_{\ell,q} \big|_{\ell_\beta=\pm \frac72} &= 6 \log(g') + \log\left(1-5I_2/(2g')^3-9 I_3/(2g')^{5}-5I_2^2/(2g')^6\right)\,,\ \dots
\end{align}
We observe that $\log \mathbb O_{\ell,q}$ vanish  for $\ell_\beta =\pm 1/2$. As we show in a moment, this property represents a nontrivial test of the
relations \re{As-beta}.

Let us examine Eq.\ \re{O-split} for $\ell=\pm 1/2$ and $\beta=0$. In this case, the symbol \re{FH} is regular and
the determinants of the corresponding Bessel operators should be described by the Szeg\H{o}-Akhiezer-Kac formula \re{SAK-B}. Indeed,  for $\beta=0$, we find from \re{O-split} that 
$\mathbb O_{\ell=\pm 1/2}=\exp(-2g I_0 +B)$ with  $B$ given by \re{1st}. We would like to emphasize, that for $\beta=0$ the octagons $\mathbb O_{\ell=\pm 1/2}$ do not receive 
corrections suppressed by powers of $1/g$. The same is true for their product $\mathbb O_{-1/2}\mathbb O_{1/2}$ which coincides with the determinant of the Wiener-Hopf operator
 \re{OO=W}. Vanishing of power suppressed corrections to the latter is exactly what one should expect because, according to the Geronimo-Case-Borodin-Okounkov formula~\cite{GC,Borodin1999AFD}, 
 the subleading corrections to the determinant of the Wiener-Hopf operator with a regular symbol are exponentially small at large $g$~\cite{BasorChen03}. For generic $\ell$, subleading corrections 
 to \re{SAK-B} are power suppressed. They can be found from \re{O-split} for $\beta=0$. For half-integer $\ell$'s, these corrections can be obtained in a closed form from \re{exa}.

\subsection{Null limit}\label{sect:null1}

Let us examine the octagon \re{O-split} in the null limit introduced in \re{null}. We demonstrated in Section~\ref{sect:null}, that, at weak coupling, the octagon simplifies in this limit. Its leading 
asymptotic behavior for $y\to\infty$ with $s=2gy$ held fixed is described by the Fredholm determinant of the Bessel operator \re{equiv} with a sharp cut-off function $\chi(x) = \theta(s-\sqrt{x})$. 
For  $s\gg 1$, it takes the form \re{SAK-weak}. At strong coupling, the octagon is given by the same determinant but with the cut-off function replaced with \re{chi-pm} and \re{chi}. The transition 
from weak to strong coupling can be interpreted as a variation of the determinant of the Bessel operator $\mathbf B_\ell(\chi)$ as a function of $\chi$.

Going to the null limit at strong coupling, we take $g> y\gg 1$ and put $\xi=0$ for simplicity. The dependence of the octagon \re{O-split} on the kinematical variables is carried by the function $B$ 
and the profile functions $I_k$. The asymptotic behavior 
of the latter was determined in \cite{Belitsky:2020qrm}
\begin{align}
I_0={y^2+\pi^2\over 2\pi} \,,\qqqquad  I_1= {\log y\over \pi}+O(y^0)\,,\qqqquad I_k=O(y^0)\,,
\end{align}
for $k\ge 2$. 
Taking into account these relations, we find that for $g'\gg 1$ or equivalently $g\gg \log y/\pi$, the expression on the right-hand side of  \re{Oq-ser} does not depend on $y$. The 
asymptotic behavior of the function $B$ is given by (see Eq.~\re{B-as})
\begin{align} 
B =
{}& \frac{ \log 2}{2\pi ^2} y^2+ \frac{\ell}2y+\frac{\log y}{4}+O(y^0)\,.
\end{align}
Substituting these relations into \re{O-split}, we get
\begin{align} \label{SAK-str}
\log \mathbb O_\ell  = -  {y^2\over \pi} \lr{g-\frac{ \log 2}{2\pi}} + \frac{\ell}2y+\frac{\log y}{4}
+c_\ell(g)+O(\log y/g)\,,
\end{align}
where $c_\ell(g)=-g \pi + (\beta\ell + \beta^2/2)\log g+ O( g^0)$ does not depend on $y$. Comparing the relations 
\re{SAK-weak} and \re{SAK-str}, we observe that $\log \mathbb O_\ell$ has the same dependence on $y$ both
at weak and strong coupling but the coefficients in front of $y^2$, $y$, $\log y$ and $O(y^0)$ terms have different dependence on the coupling constant and the bridge length $\ell$.

\subsection{Large bridge limit}\label{sect:large}

Deriving the strong coupling expansion in the previous section, we tacitly assumed that the bridge length $\ell$ stays
finite as $g\to\infty$. In this subsection, we examine the limit when $\ell$ scales as $O(g)$  with $\bar \ell=\ell/(2g)$ held fixed. 
This limit was previously studied in Ref.~\cite{Bargheer:2019exp}. It was shown there, that 
the leading term $A_0$ in the expansion of the octagon \re{O-str} acquires a nontrivial dependence on $\bar\ell$ (see Eq.~\re{i0} below). In this section, 
we apply the results obtained above to describe dependence of the octagon on $\bar\ell$.

We recall, that the expansion coefficients in \re{As-beta} are polynomial in $\ell_\beta=\ell+\beta$. At large $\ell$, we have $A_{2k}= O(\ell^{2k})$ and $A_{2k+1}= O(\ell^{2k})$. As a 
consequence, for $\ell=O(g)$, the contribution to \re{O-str} from terms with even and odd coefficients scales as $O(g)$ and $O(g^0)$, respectively. Neglecting the latter, we examine 
the asymptotic behavior of the even coefficients in \re{As-beta} at large $\ell$
\begin{align}\label{A-large} 
& A_2=-I_1\ell^2\, ,  && A_4=-\frac{I_2}{8}\ell^4\, ,  && A_6=-\frac{3
   I_3}{128}\ell^6\,,  &&  A_8=-\frac{5 I_4}{1024}\ell^8\,, \ \dots
\end{align}
A close examination reveals that the rational factors in \re{A-large} coincide with the expansion  coefficients of
$-2x^2/\sqrt{4-x^2}$ at small $x$. This leads to the following integral representation for $A_{2k}$ with $k\ge 1$
\begin{align}\label{A-int}
A_{2k} 
   =  - \int_C {dx\over 2\pi i} {{2x^2\over \sqrt{4-x^2}} }{\ell^{2k}\over x^{2k+1}} I_k\,,
\end{align}
where the integration contour $C$ encircles two cuts on the real axis located at $x\le -2$ and $x\ge 2$. 

To find the contribution of \re{A-int} to the strong coupling expansion  \re{O-str}, we introduce the generating function
\begin{align}\label{gen}
F(\bar\ell) = \sum_{k\ge 1} {A_{2k}\over g^{2k}}  =  -\int_C {dx\over 2\pi i} {{2x\over \sqrt{4-x^2}} }\int_0^\infty {dz\over\pi} z\partial_z \log(1-\widehat \chi(z))
{4\bar\ell^2 \over x^2 z^2-4\bar\ell^2 }\,,
\end{align}
where $\bar\ell=\ell/(2g)$. Here we replaced the profile function $I_k$ with its integral representation \re{In} and performed the summation over $k$. The integrand in the last 
relation has poles at $x=\pm 2\bar\ell/z$. For $z^2 < \bar\ell^2$ these poles are located on the branch cuts
$x^2>4$ and, deforming the integration contour $C$ away from the cuts, we find that they do not contribute to
$F(\bar\ell)$. For $z^2 > \bar\ell^2$, we pick up residues at the poles $x=\pm 2\bar\ell/z$ to get
\begin{align} \label{sum-A}
F(\bar\ell) = -{4\over\pi}\int_{\bar\ell}^\infty {dz\over z} \partial_z \log(1-\widehat \chi(z)){\bar\ell^2\over \sqrt{1-\bar\ell^2/z^2}} 
+  4\beta\bar\ell\,.
\end{align}
The additional term in the right-hand side comes from the residue at infinity. It ensures that  
$F(\bar\ell) =O(\bar\ell^2)$ for small $\bar\ell$.
 
Using the generating function \re{gen}, we can evaluate \re{O-str} as
\begin{align}\label{O-ll}
\log \mathbb O_\ell = -g \bigg[ A_0 + 2 \beta\bar\ell(\log(2\bar \ell )-1) + {\bar\ell}\log b(0)  - \frac{1}2
\int_0^1{ds\over s^2}(1-s) F(\bar\ell s)\bigg] + O(g^0)\,,
\end{align}
where we replaced $A_1$ and $B$ defined in \re{uni} by their leading asymptotic behavior at large $g$. The last term in the right-hand side of \re{O-ll} describes the contribution 
of terms with odd coefficients $A_{2k+1}$. Substituting \re{sum-A} into \re{O-ll}, we find after some algebra that the expression inside
the brackets in \re{O-ll} takes a simple form
\begin{align}\label{O-bar}
\log\mathbb O_\ell(\widehat\chi) =  {2g} \int_{\bar\ell}^\infty {dz\over\pi}  \log(1-\widehat \chi(z))  \sqrt{1-\bar \ell^2/z^2} + O(g^0)\,,
\end{align}
where $\bar\ell=\ell/(2g)$.
We verify, that for $\bar\ell=0$ this expression coincides with the leading term of the strong coupling expansion \re{O-str}.
For nonzero $\bar\ell$, the relation \re{O-bar} takes into account corrections to the octagon of the form $g(\ell/g)^{k}$ to all orders in $k$. Using expressions for the expansion 
coefficients \re{As-beta}, it should be possible to determine subleading corrections to \re{O-bar}. This question deserves further investigation.

From the point of view of the strong Szeg\H{o} limit theorem, the relation \re{O-bar} provides the leading asymptotic behavior of the  Fredholm determinant of the Bessel operator 
$\mathbf B_\ell(\widehat\chi)$ in the double scaling limit of large $g$ and $\ell$ with their ratio held fixed.

Another representation for the `simplest' octagon $\mathbb O_\ell^{(s)}$ with a large bridge length $\ell=O(g)$ was derived in Ref.~\cite{Bargheer:2019exp} using the clustering procedure~\cite{Jiang:2016ulr} 
\begin{align}\label{i0}
\log \mathbb O_\ell^{(s)} = {g \over\pi}\int_{-\infty}^\infty d\vartheta \lr{\xi \cosh\vartheta + i \bar\ell\, {\cosh^2\vartheta\over\sinh(\vartheta-i0 \xi)}} \log(1+Y(\vartheta))\,,
\end{align}
where the function $Y(\vartheta)$ is defined as
\begin{align}\label{Y}
Y(\vartheta) =
- \frac{\cosh y+ \cosh \xi}{\cosh y+ \cosh ( \xi \cosh \vartheta + i \bar\ell \sinh \vartheta ) }\,.
\end{align}
The second term inside the brackets in \re{i0} has a pole at $\vartheta=0$, the integration contour  is deformed using the $\pm i0$ prescription, depending on the sign of $\xi$.

To show the equivalence of the two representations \re{O-bar} and \re{i0}, we change the integration variable in \re{i0} from the real $\vartheta$ to a complex $z$
\begin{align}\label{new-var} 
& z^2 =(\xi^2 -\bar\ell^2 )\sinh^2\vartheta +2i \bar\ell \xi \sinh\vartheta\cosh\vartheta \,.
\end{align}
The integration contour $C_z$ in the complex $z-$plane has two branches $z_+$ and $z_-$ corresponding to $\vartheta>0$ and $\vartheta<0$, respectively. They are complex conjugate to each 
other, $z_-=z_+^*$ with ${\rm Im} \,z_+>0$, and merge at the origin. We have $z_+\sim (2i \bar\ell \xi \vartheta)^{1/2}$ for $\vartheta\to 0$  and $z_+\sim \e^\vartheta(\xi+ i\bar\ell)/2$ for $\vartheta\to\infty$. 

Changing the variables to \re{new-var}, we get from \re{i0}
\begin{align}\label{Cz}
\log \mathbb O_\ell {}& = {g\over\pi}\int_{C_z} dz\,  \log(1-\widehat\chi(z)) \left[ {i\over z} \sqrt{\bar\ell^2-z^2}
  +  \frac{i \xi \bar\ell}{z \sqrt{\xi ^2+z^2}}\right]\,,
\end{align}
where $\widehat \chi(z)= -Y(\vartheta)$ coincides with the function $\widehat\chi^{(s)}$ defined in \re{hatchi}.
The expression inside the brackets has a pole at $z=0$ and the two branching points $z^2=-\xi^2$ and $z^2=\bar\ell^2$
which are located on different sides of the integration contour $C_z$. The prescription $(-i0 \xi)$ in \re{i0} 
transates to $z_\pm^2\sim 2i \bar\ell \xi (\vartheta-i0 \xi)$ for small $\theta$. This amounts to shifting the integration contour  in \re{Cz} around $z=0$ in such a way that $C_z$ can be further 
deformed to encircle the cut that starts at $z^2=\bar\ell^2$ and goes to infinity. Then, the second term inside the brackets in \re{Cz} gives vanishing contribution and the first one yields \re{O-bar}.
 
\section{Conclusions}\label{sect:conc}

In this paper, we developed a new technique for computing correlation functions of heavy half-BPS operators in planar $\mathcal N=4$ SYM theory for arbitrary coupling constant. The starting 
point of our analysis was a representation of these correlation functions in terms of the octagons $\mathbb O_\ell$ with an arbitrary bridge length $\ell$. We argued that $\mathbb O_\ell$ can be 
expressed as the Fredholm determinant of the integrable Bessel operator and demonstrated that this representation 
is very efficient in deriving expansion of the octagon both at weak and strong coupling. 
The presented results generalize those obtained in Ref.~\cite{Belitsky:2020qrm} for zero-length bridge. 

The introduction of a nonvanishing bridge $\ell$ leads to interesting novel features of the octagon. 
At weak coupling, perturbative corrections are enhanced in the null limit $y\to\infty$ when the four operators sit at the vertices of a null rectangle. For $\ell=0$ such corrections exponentiate 
and scale as $\log \mathbb{O}_0 \sim y^2$ to any order in $g^2$. For $\ell>0$ the corrections to
$\log \mathbb{O}_\ell$ scale as $(g^2 y^2)^{\ell + n + 1}$ with $n\ge 0$, so that  the power of $y^2$ increases with the loop order.
We found that the resummed expression for $\mathbb{O}_\ell$ is related to  the $\tau-$function of the Okamoto's theory of the Painlev\'e V equation and obeys the well-known 
Toda lattice equation.

At strong coupling, we exploited the determinant representation of the octagon to derive the first few terms of its expansion at large $g$ from the strong 
Szeg\H{o} limit theorem. To achieve this goal, we had to generalize results previously 
obtained in mathematical literature for asymptotic behavior of the determinant of the Bessel operator. As a byproduct of our analysis we formulated a (conjectured) modified
Szeg\H{o}-Akhiezer-Kac formula for the determinant of the
Bessel operator with a  Fisher-Hartwig singularity.~\footnote{We checked this formula in different ways and left its rigorous proof to specialists.  
} 
 
One of the consequences of our consideration was the elucidation of the universal origin of the coefficient $A_1$ accompanying the $\log g$ term in \re{O0}. We showed that $A_1$ 
is a linear function of the bridge length $\ell$, independent of the kinematical variables. 
This property is a manifestation of the  Fisher-Hartwig singularity and it is related to the behavior of the symbol of the Bessel operator around the origin. 

The Szeg\H{o}-Akhiezer-Kac formula describes the first three terms of the strong coupling expansion of the octagon \re{O0} and yields the coefficients $A_0$, $A_1$ and $B$. In our 
analysis, we used their expressions as initial conditions for the method of differential
equations. This allowed us to compute the coefficients of the strong coupling expansion of the octagon to any order in $1/g$. In terms of the strong Szeg\H{o} limit theorem, they determine 
subleading, power suppressed corrections to the determinant of the truncated Bessel operators and closely related Wiener-Hopf operators. We found that these corrections have an interesting 
iterative structure and can be resummed to all orders in $1/g$ for specific values of the bridge length $\ell$.
  
There is a number of important issues that deserve further investigation.  Analyzing subleading corrections to the octagon at strong 
coupling, we systematically ignored exponentially small terms. The latter are interesting in their own right as they may shed some light on non-perturbative effects
at strong coupling. For the Wiener-Hopf operators with regular symbols, the Geronimo-Case-Borodin-Okounkov formula provides a systematic way
to account for exponential suppressed terms. It would be interesting to find its analogue for the Bessel operators with regular and singular symbols, which in the 
present situation describe `asymptotic' and `simplest' octagons, respectively.

When discussing the octagons with large bridge length $\ell=O(g)$, we found that the leading terms of the form $g(\ell/g)^n$ can be resummed to all orders into a simple expression 
that matches result previously obtained using the clustering technique. It would be interesting to extend this analysis to subleading effects and find a generalization of the Szeg\H{o}-Akhiezer-Kac
formula in the double scaling limit $g, \ell \to \infty$ with $\bar\ell=\ell/(2g)$ held fixed.

The formalism developed in this paper can be applied to computing other observables in $\mathcal N=4$ SYM which are related to operator determinants. One such example of particular
interest is the behavior of the six-point gluon MHV scattering amplitude at a specific kinematical point corresponding 
to zero values of some Mandelstam invariants~\cite{Basso:2020xts}. The leading double logarithmic behavior of the amplitude is controlled by 
anomalous dimensions whereas  the constant term can be cast as a determinant of the tilted flux-tube kernel.
At strong coupling, it has the leading asymptotic behavior similar to that of the octagon \re{O0}. The reason for this is that 
the twisted flux-tube kernel defines an integral operator with a matrix symbol possessing a Fisher-Hartwig singularity and its determinant can be computed by
the technique developed in this work.  

\section*{Acknowledgments}

We would like to thank Ivan Kostov and Didina Serban for interesting discussions and Riccardo Guida for his help with numerical calculations. The research of A.B.\  and  G.K.\  was supported, 
respectively, by the U.S. National Science Foundation under the grant PHY-1713125 and by the French National Agency for Research grant ANR-17-CE31-0001-01.

\appendix

\section{Bessel operator with Fisher-Hartwig singularity}\label{app:HF} 
 
In this appendix, we formulate a conjecture \re{SAK} for the Fredholm determinant of the Bessel operator \re{B-hat} with the symbol having a  Fisher-Hartwig singularity of the form \re{FH}. 
 
According to its definition \re{B-hat}, the truncated Bessel operator $\mathbf B_\ell (\widehat \chi)$ acts on the interval $[0,2g]$ and depends on the symbol $\widehat\chi(z)$. For sufficiently 
smooth function $\widehat \chi(z)$ verifying the condition $\widehat \chi(z) \neq 1$ for $0\le z<\infty$, the Fredholm determinant of $\mathbf B_\ell (\widehat \chi)$ is given 
at large $g$ by the relation \re{SAK-B}, see Ref.~\cite{BasorEhrhardt03}. 

The relation \re{SAK-B} is a generalization of the strong Szeg\H{o} limit theorem to the truncated Bessel operator. This theorem has been originally formulated for determinants of 
Toeplitz matrices of large size~\cite{Szego:1915,Szego:1952}. In continuum limit, the latter become the determinants of truncated Wiener-Hopf operators $\mathbf W(\chi)$ defined in 
\re{WH}.~\footnote{The symbol of the operator \re{WH} is an even function of $z$, this condition can be relaxed for a generic definition of the  Wiener-Hopf  operator. } As in the case of the 
Bessel operator, it is convenient to switch to the function $\widehat\chi(z) = \chi(4g^2 z^2)$ and use an equivalent representation of the same operator
\begin{align}\notag\label{WH-hat}
& \mathbf W(\widehat\chi) f(t) = \int_{-2g}^{2g} dt'\,\widehat W(t-t') f(t')\,,
\\
& \widehat W(t-t') = \int_{0}^\infty {dz\over\pi}\,\cos(z(t-t')) \widehat\chi(z)\,.
\end{align}
The operator $\mathbf W(\widehat\chi)$ acts on the interval $[-2g,2g]$ and, in virtue of translation invariance of the kernel, its determinant only depends on the length of the interval.

The strong Szeg\H{o} limit theorem states that the Fredholm determinants of the Wiener-Hopf and Bessel operators with a regular symbol $\widehat\chi$ are given at large $g$ by~\cite{Akhiezer:1964,Kac:1964,BasorEhrhardt03}  
\begin{align}\label{det-W}\notag
& \det (1-\mathbf W(\widehat\chi))_{[-2g,2g]} = \exp\lr{4g\widetilde \psi(0) + \int_0^\infty dk\,k \big(\widetilde \psi(k)\big)^2 + \dots }\,,
\\
& \det\big(1-{\mathbf B}_\ell(\widehat{\chi})\big)_{[0,2g]}
= \exp \bigg( 2g \widetilde \psi(0) - {\ell\over 2} \psi(0) + \frac12 \int_0^\infty dk\, k \big(\widetilde\psi(k)\big)^2 + \dots\bigg)\,,
\end{align}
where $\widetilde \psi(k)$ is defined in \re{psi}. Here the ellipses denote corrections vanishing as $g\to\infty$. 
According to the Geronimo-Case-Borodin-Okounkov formula, the subleading corrections to the first relation in \re{det-W} are exponentially small at large $g$~\cite{BasorChen03}. For the 
Bessel operator, analogous formula is not known in the literature. In fact, we show in Section~\ref{sect:exp}, that subleading corrections to the second relation in \re{det-W} are instead power 
suppressed at large $g$ . We also develop there a technique that allows us to compute these corrections to any order in $1/g$, see Eq.~\re{Delta} below.
 
For the singular symbol $\widehat\chi$, the relation \re{det-W} has to be modified. 
In our analysis, we encountered singular symbols of two different kinds. 
The first one is
$\chi(x) = \theta(s^2-x)$ or equivalently $\widehat\chi(z) = \theta(s^2-4g^2 z^2)$. We have shown in Section~\ref{sect:Toda} that this function describes the asymptotic behavior of the octagon 
at {\it weak} coupling  in the null limit. In this case, the Fredholm determinants of the Wiener-Hopf and Bessel operators are given by~\cite{Mehta,Tracy:1993xj,EHRHARDT20103088}
\begin{align}\notag\label{sharp}
& \det (1-\mathbf W( \chi))= \exp\lr{-{s^2\over 8}-\frac14 \log s  +\frac1{3}\log 2 + 3 \zeta'(-1) + O(1/s) }\,,
\\
& \det (1- \mathbf B_\ell( \chi)) = \exp\lr{ -{s^2\over 4} +\ell s -{\ell^2\over 2} \log s + \log{G(\ell+1)\over (2\pi)^{\ell/2}}+ O(1/s) }\,.
\end{align}
where $\zeta'(-1)={1}/{12}-\log A$ is the derivative of the Riemann zeta function and $A$ is the Glaisher constant. 
Notice that the exponents in both relations scale quadratically with $s$ and subleading corrections  run in powers of $1/s$. The second relation in \re{sharp} leads to \re{SAK-weak}.

The second singular symbol takes the form \re{FH}. It has the Fisher-Hartwig singularity and describes the asymptotic behavior of the `simplest' octagon at {\it strong} coupling. 
For the symbol \re{FH}, the Fredholm determinant of the Wiener-Hopf operator \re{WH-hat}
was derived so far only for specific values of $\beta$ in Ref.~\cite{Bottcher89} and its general expression was conjectured to be, see, e.g., Ref.~\cite{Bttcher2006AnalysisOT},
\begin{align}\notag\label{pre1}
 \det (1-\mathbf W(\widehat\chi))_{[-2g,2g]}= \exp\bigg[ 4g \widetilde \psi(0) + \int_0^\infty dk\, \left( k(\widetilde \psi(k))^2-\beta^2 {1-\e^{-k}\over k}\right)&
\\
+\beta^2 \log(2g) + \log {G^2(\beta+1)\over G(2\beta+1)} + O(1/g)\bigg]& \,,
\end{align}
where the function $\widetilde \psi(k)$ is defined in \re{psi} and $G$ is the Barnes function. 
As compared with \re{det-W}, this relation contains the additional terms generated by the Fisher-Hartwig singularity.
It follows from \re{FH} and \re{psi} that $\widetilde\psi(k) \sim -\beta/k$ at large $k$. The $\beta-$dependent term 
on the first line of \re{pre1} is needed for the integral to be well-defined. 

For the determinant of the Bessel operator with the symbol \re{FH} we put forward the following conjecture
\begin{align}\notag\label{SAK1}
\det\big(1-\mathbf B_\ell(\widehat\chi)\big)_{[0,2g]} {}& = \exp\bigg[ 2g \widetilde \psi(0)   -\frac{\ell}2\log b(0)+ \frac12 \int_0^\infty dk\, \left( k(\widetilde \psi(k))^2-\beta^2 {1-\e^{-k}\over k}\right)
\\
{}&+
\lr{\beta\ell + \frac12\beta^2}\log g
+ \frac{\beta}2   \log(2\pi)  + \log {G(1+\ell)\over G(1+\ell+\beta)} + O(1/g)\bigg],
\end{align}
where the function $b(z)$ is defined in \re{FH}. We show in Section~\ref{app:B} that for $\ell=0$ and $\widehat\chi=\widehat\chi^{(s)}$ (see Eq.~\re{hatchi}) the relation \re{SAK1} correctly 
reproduces numerical results for the determinant of the Bessel operator obtained in Ref.~\cite{Belitsky:2020qrm}. In addition, the relation \re{SAK1} can be checked in various ways. 

We recall that the determinants of the Wiener-Hopf and Bessel operators are related to each other by the relation \re{sharp}.
It is straightforward to verify that the two expressions in \re{sharp} satisfy \re{BB-W}. Analogous relation holds 
between determinants of the operators $\mathbf W(\widehat\chi)$ and $\mathbf B_{\pm 1/2}(\widehat\chi)$
\begin{align}\label{BBhat}
\det\big(1-{\mathbf B}_{1/2}(\widehat{\chi})\big)_{[0,2g]}\det\big(1-{\mathbf B}_{-1/2}(\widehat{\chi})\big)_{[0,2g]}= \det (1-\mathbf W(\widehat\chi))_{[-2g,2g]}\,.
\end{align}
We verify using \re{det-W} that this relation holds for a regular symbol $\widehat\chi$. Substituting \re{pre1} and \re{SAK1} 
into \re{BBhat} we find that it is also satisfied for the singular symbol \re{FH}.
 
Choosing $b(z)=1$ in \re{FH} and applying \re{SAK1}, we can obtain the Fredholm determinant for the Bessel operator with the symbol
$\widehat\chi(z)=1-\omega_\beta(z)$. The calculation shows that
\begin{align}\label{use}
\int_0^\infty {dz\over\pi} \log \omega_\beta(z) \cos(z x) = -{\beta\over k}\lr{1-\e^{-k}}\,,
\end{align}
where $\omega_\beta(z)$ is defined in \re{FH}.
Replacing $\widetilde\psi(k)$ in \re{SAK1}  with this expression,  we find after some algebra 
\begin{align}\notag
\det\big(1-\mathbf B_\ell(1-\omega_\beta)\big)_{[0,2g]} {}& = \exp\bigg[ -2g \beta+\lr{\beta\ell + \frac12\beta^2}\log g  
\\[1.5mm]
{}& +\frac12\beta\log(2\pi) -\frac12 \beta^2 \log 2 + \log {G(1+\ell)\over G(1+\ell+\beta)} +O(1/g)\bigg]\,.
\end{align}
For $\ell=\pm 1/2$, this relation agrees with the findings of Ref.~\cite{BasorEhrhardt05}.  
 
The leading terms in Eqs.\ \re{pre1} and \re{SAK1} are proportional to $\widetilde\psi(0) = \int_0^\infty {dz}  \log(1-\widehat\chi(z))/\pi$. Replacing $\widehat\chi(z)$ with its expression \re{FH}, we get
\begin{align}
\widetilde\psi(0) = \int_0^\infty {dz\over\pi}  \log b(z) +\int_0^\infty {dz\over\pi}   \log \omega_\beta(z)\,.
\end{align}
The two terms in the right-hand side define the leading asymptotics of the determinants of the operators with symbols
$1-b(z)$ and $1-\omega_\beta(z)$, respectively. This suggest that the appropriately taken ratio of the determinants should approach a finite
value for $g\to\infty$. For the Wiener-Hopf operator $\mathbf W(\widehat\chi)$, this is known as the localization property \cite{Basor79}
\begin{align}\label{loc}
\lim_{g\to\infty}{\det(1-
\mathbf W(\widehat\chi))\over \det (1-\mathbf W(1-b)) \det (1-\mathbf W(1-\omega_\beta))} =  \e^{2C}\,,
\end{align}
where $C$ does not depend on $g$ and is defined below. 

We can show using \re{SAK1} that the determinant of the Bessel kernel $\mathbf B_\ell(\widehat\chi)$ has the same property 
\begin{align}\label{loc1}
\lim_{g\to\infty}{\det(1-\mathbf B_\ell(\widehat\chi))\over \det (1-\mathbf B_\ell(1-b)) \det (1-\mathbf B_\ell(1-\omega_\beta))} =  \e^{C}\,,
\end{align}
where the constant $C$ is the same as in \re{loc}. It is given by
\begin{align}\notag
C {}& =\int_0^\infty dk \, k \int_0^\infty {dz\over\pi} \, \log b(z)\cos(zk) \int_0^\infty {dz'\over\pi} \, \log \omega_\beta(z') \cos(z'k)  
\\ &
=-\beta \left( \frac12 \log  b(0)- \int_{-\infty}^\infty {dz\over 2\pi} \,  {\log b(z)\over 1+z^2}\right).
\end{align} 
Here we applied \re{use} and used the parity property  $b(z)=b(-z)$ to extend the integration region to the whole real axis.~\footnote{For the symbol $\widehat\chi^{(s)}$ defined in \re{hatchi}, $b(z)$ 
is an even function of $z$. For arbitrary $b(z)$ the expression for $C$ is more involved. }
The integral in the last relation can be evaluated by applying the Wiener-Hopf decomposition $b(z) = b_+(z) b_-(z) = b_+(z) b_+(-z)$ where $b_+$ and $b_-$ are analytical in the upper and 
lower half-plane, respectively. Then, computing the integral by residues, we get
\begin{align}
C 
= \beta \log {b_+(i)\over b_+(0)}.
\end{align}
Notice that $C$ is linear in $\beta$ and it is independent of the index $\ell$ of the Bessel function. 
Universality of the constant in the right-hand side of \re{loc} and \re{loc1} follows from \re{BBhat}.

The Szeg\H{o}-Akhiezer-Kac relations \re{pre1} and \re{SAK1} hold up to contributions suppressed by powers of $1/g$.
Denoting the $O(1/g)$ corrections to the exponents of \re{pre1} and \re{SAK1} as $\Delta_W$  and $\Delta_{B,\ell}$, 
respectively, we find from \re{O-str} 
\begin{align}\label{Delta}
\Delta_{B,\,\ell}= \sum_{k\ge 2} {A_k\over 2k(k-1)} g^{1-k}\,,\qqqquad 
\Delta_{W} = \Delta_{B,\,-1/2}+ \Delta_{B,\,1/2}\,,
\end{align}
where the expansion coefficients are given by \re{As-beta} and the second relation follows from \re{BBhat}.

We demonstrated in Section~\ref{sect:exp}, that power suppressed corrections to the determinant of the Bessel operator have an interesting iterative form and they can be found in a closed 
form for specific values of parameters, see, e.g., Eqs.~\re{exa} and \re{O-bar}. For the Wiener-Hopf operator, we find from \re{Delta} and \re{As-beta}
\begin{align}\notag
\Delta_W =& -\frac{\beta ^2 I_1}{2 g}-\frac{\beta ^2 I_1^2}{8 g^2}-\frac{\beta ^2
   \left(\left(\beta ^2-1\right) I_2+4 I_1^3\right)}{96 g^3}-\frac{\beta
   ^2 \left(\left(\beta ^2-1\right)I_1 I_2+I_1^4\right)}{64
   g^4}
\\   
  &-\frac{\beta ^2 \left(20  (\beta ^2-1) I_2
   I_1^2+ (\beta ^2-1) (\beta ^2-4) I_3+8
   I_1^5\right)}{1280 g^5}+O(1/g^6)\,.
\end{align}
As expected, $\Delta_W$ vanishes for $\beta=0$. Using \re{exa}, we deduce  
\begin{align}\notag
\Delta_{W} (\beta=\pm 1)& =\log\left(g'/g\right),
\\[2mm]\notag
\Delta_{W} (\beta=\pm 2)&=4\log(g'/g) + \log\left(1-I_2/(2g')^3\right),
\\[2mm]\notag
\Delta_{W} (\beta=\pm 3)&=9\log(g'/g) + \log(1-I_2/(2g')^3)
\\[2mm]
&\phantom{=}
+\log\left(1-5I_2/(2g')^3-9 I_3/(2g')^{5}-5I_2^2/(2g')^6\right),
\end{align}
where $g'=g-I_1/2$ and the functions $I_k$ are introduced in the main text in Eqs.\ \re{In} and \re{Ibeta}.

\section{Application of  Szeg\H{o}-Akhiezer-Kac formula } \label{app:B}

In this appendix, we apply the Szeg\H{o}-Akhiezer-Kac formula \re{SAK}  to compute the function $B^{(s)}$ 
defined in \re{2nd} 
in various kinematical limits and compare the obtained results 
with the findings of Ref.~\cite{Belitsky:2020qrm}.  

It is convenient to rewrite the expression for  $B_\ell(y,\xi)\equiv B^{(s)}$ in \re{2nd} as
\begin{align}\notag\label{BB}
 {}& B_\ell(y,\xi) = B_0(y,\xi) - \log \Gamma(1+\ell) -\frac12 \ell\log \lr{\frac{\sinh \xi}{2 \xi  (\cosh \xi+\cosh y)}}\,,
 \\\notag
 {}&  B_0(y,\xi) = \frac12 \log (2\pi)  + \frac12 \int_0^{\infty} {dk} \left( k (\widetilde\psi(k))^2
-{1-\e^{-k}\over k}\right)\,,
\\ 
 {}& \widetilde\psi(k) = \int_0^\infty {dz\over\pi} \, \log\left({\cosh\sqrt{z^2+\xi^2}-\cosh\xi\over \cosh\sqrt{z^2+\xi^2}+\cosh y}\right)\cos(k z)\,.
 \end{align}
The variables $y$ and $\xi$ parameterize  the cross-ratios \re{y-xi} and \re{x-ratio}.
We consider four different kinematical regions, we refer the reader to Ref.~\cite{Belitsky:2020qrm} for explanation of their physical significance.
 
\subsection*{$\boldsymbol{ y=\xi=0}$ }

In this case, we have
\begin{align}\label{sim1}  
&\widetilde \psi(k)= \int_{0}^\infty {dz\over \pi} \log(\tanh^2(z/2))\cos(zk) 
= - {1\over k} \tanh(k\pi/2)\,.
\end{align}
Notice that $\psi(k)\sim -1/k$ at large $k$. As explained in Section~\ref{sect:FH}, this is a manifestation of the Fisher-Hartwig singularity. The integral in the expression for $B_0$ in \re{BB}  
is well-defined and its calculation yields
\begin{align}
B_{0}= -6 \log A +\frac{1}{2}+\frac{7 \log 2}{6}+\log \pi = 0.960875\,,
\end{align}
where $A$ is the Glaisher's constant.
This result is very close to the numerical value  found in  Ref.~\cite{Belitsky:2020qrm},  see Eq.\ (6.9) there. For arbitrary bridge length, we have
\begin{align}
B_\ell=- \log \Gamma(1+\ell) + \ell\log 2 - 6 \log A+\frac12+\frac{7 \log 2}{6}+ \log \pi \,.
\end{align}

\subsection*{$\boldsymbol{\xi=0,\ y=i(\pi-\phi)}$}

The calculation of $\widetilde\psi(k)$ gives
\begin{align} \label{psi-k} 
& \widetilde\psi(k) = \int_{0}^\infty {dz\over \pi}\log \left(\frac{\cosh z-1}{\cosh z-\cos \phi}\right)\cos(kz)
= {\cosh ((\pi-\phi) k)-\cosh(\pi k)\over k\sinh(\pi k) }\,.
\end{align}
At small $\phi$, we get from \re{BB}
\begin{align}\label{Bl}
B_\ell =  -\log \Gamma(1+\ell) + \ell\log \phi + \frac12 \log(\phi \pi) + \dots\,,
\end{align}
where dotes denote terms vanishing as $\phi\to 0$. For $\ell=0$ this relation coincides with the analogous expression in 
 Ref.~\cite{Belitsky:2020qrm}, see Eq.~(6.17). 

\subsection*{$\boldsymbol{\xi=0,\ y\gg 1}$ }

Taking into account \re{psi-k}, we get
\begin{align} 
& \widetilde\psi(k) = \int_{0}^\infty {dz\over \pi}\log \left(\frac{\cosh z-1}{\cosh z+\cosh y}\right)\cos(kz)
=  {\cos (y k)-\cosh(\pi k)\over k\sinh(\pi k) }\,.
\end{align}
Substituting this expression into \re{BB}, we obtain after some algebra
\begin{align} \notag
B_0(y,\xi) &= \frac12 \log (2\pi) 
+  \int_0^\infty {dk\over 2k } \left({\e^{-k} - {1\over \cosh^2(\pi k/2)}} \right)
\\
&+ \int_0^\infty {dk\over 4k } 
\bigg[
{(\cos(yk)-1)^2\over \cosh(\pi k)-1}+  {4-(\cos(yk)+1)^2\over \cosh(\pi k)+1}\bigg].
\end{align}
The integral in the second line can be evaluated at large $y$ by replacing $1/(\cosh(\pi k)\pm 1)$ with their  Mellin-Barnes representation. 
This leads to the following result for $B_\ell$ at large $y$
\begin{align}\label{B-as} 
B_\ell =
{}& \frac{ \log 2}{2\pi ^2} y^2+ \frac{\ell}2y+\frac{\log y}{4}-\log{\Gamma(1+\ell)}  -3 \log A+\frac{1}{4} +\frac{5\log 2}{6} + \frac34\log \pi +O(1/y^2)\,.
\end{align}
For $\ell=0$ this relation agrees with with eq.(6.27) in \cite{Belitsky:2020qrm}.
 
\section*{$\boldsymbol{y=i\pi,\ \xi\gg 1}$ }

In this limit, the expression for $\widetilde\psi(k)$ can be simplified as
\begin{align}
\widetilde\psi(k) \sim \int_{0}^\infty {dz\over \pi} \log(1-\e^{-z^2/(2\xi)})\cos(k z) = -\lr{\xi\over 2\pi}^{1/2} \sum_{p\ge 1}p^{-3/2} \e^{-k^2\xi/(2p)}\,,
\end{align}
where in the second relation we replaced logarithm by its series expansion. Replacing $\widetilde\psi(k)$ in \re{BB}
with this expression, we obtain a representation of $B_\ell$ as a double series. Its resummation gives
\begin{align}
B_\ell {}& =\frac14 \lr{2\ell+1} \log (2 \xi )- \log \Gamma(1+\ell) + \frac14\log \pi + c + \dots\,,
\end{align}
where dots denote terms vanishing for $\xi\to\infty$ and a notation was introduced for
\begin{align}
c=\frac14\log(4\pi)+ {1\over 4\pi}\sum_{p\ge 0} (-1)^p \zeta \left(\ft{1}{2}-p\right) \zeta \left(p+\ft{3}{2}\right) =  0.348761\,.
\end{align}
For $\ell=0$ the expression for $B_0$ agrees with Eq.~(6.12) in \cite{Belitsky:2020qrm}.

\section{Relation between Bessel and Wiener-Hopf operators} \label{app:WH}

In this Appendix, we follow Ref.~\cite{Mehta} to prove the relation \re{OO=W}.  
For half-integer $\ell$, the Bessel functions $J_\ell(x)$ are expressible in terms of trigonometric functions, e.g.,
$J_{-1/2}(x) = \lr{2/(\pi x)}^{1/2} \cos x$ and $J_{1/2}(x) = \lr{2/( \pi x)}^{1/2}\sin x$.
Using these relations, we get from \re{B-ker}
\begin{align}\label{B-simp}
B_{\mp 1/2}\left(t_1^2,t_2^2\right) = \frac12 (t_1 t_2)^{-1/2}\left[ W(t_1-t_2) \pm W(t_1+t_2)\right]\,,
\end{align}
where the function $W(t_1-t_2)$ is the kernel of the Wiener-Hopf operator \re{WH}. In a similar manner, the function
$W(t_1+t_2)$ defines the kernel of the Hankel operator. Equation \re{B-simp} implies that for $\ell=\pm 1/2$ the Bessel operator
reduces to a sum or difference of truncated Wiener-Hopf and Hankel operators acting on the interval $[0,1]$, see, e.g., Ref.~\cite{basor2003determinant}.

Let us examine the eigenproblem for the operator $\mathbf W(\chi)$ defined in \re{WH}
\begin{align}
 \int_{-1}^1 dt'\,W(t-t') f_\lambda(t') =\lambda f_\lambda(t)\,.
\end{align}
Since $W(t)$ is an even function of $t$, the eigenfunctions $f_\lambda(t)$ have a definite parity under $t\to -t$.
Taking into account \re{B-simp}, we find that the even and odd eigenfunctions, $f_\lambda^{(+)}$ and $f_\lambda^{(-)}$, respectively, satisfy the equation
\begin{align}
2\int_0^1 dt' \, (t\,t')^{1/2}B_{\mp 1/2}\left(t^2,t'{}^2\right)f_\lambda^{(\pm)}(t') = \lambda^{(\pm)}f_\lambda^{(\pm)}(t)\,.
\end{align}
Next, changing the integration variables to $t^2$ and ${t'}^2$, we observe that $f_\lambda^{(\pm)}(\sqrt{t})/t^{1/4}$ diagonalize the operators $ \mathbf B_{\mp 1/2}(\chi)$ introduced in \re{B-ker}. 
As a consequence, $\det(1-\mathbf B_{\mp 1/2}(\chi)=\prod \lambda^{(\pm)}$ leads to
\begin{align}\label{BB-W}
\det\big(1-\mathbf B_{-1/2}(\chi)\big)\det\big(1-\mathbf B_{1/2}(\chi)\big)=\det(1-\mathbf W(\chi))\,.
\end{align}
Combining this relation with \re{equiv}, we finally arrive at \re{OO=W}.
 
\section{Numerical tests}\label{app:num}
 
In this appendix, we test the obtained expressions for the octagon by comparing them with the results of numerical evaluation performed using the technique described in Ref.~\cite{Belitsky:2020qrm}.

\subsection*{Null limit}

\begin{figure}[h!t]
\psfrag{g}[cc][cc]{$g$}\psfrag{LnO}[cc][cc]{$(\log \mathbb O_1)/g$}
\centerline{\insertfig{10}{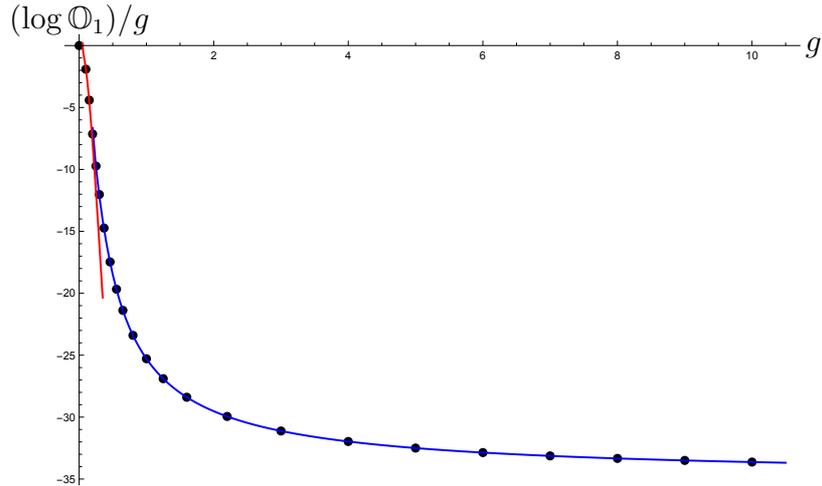}}
\caption{The dependence of the octagon $\mathbb O_{\ell=1}(\widehat\chi^{(s)})$ on the coupling constant for $y=10$ and $\xi=0$. Black dots denote numerical values, 
red and blue curves describe weak and strong coupling expansions, respectively. 
}\label{fig:null}
\end{figure}

Let us consider the `simplest' octagon \re{equiv} with the symbol given by the function $\widehat\chi^{(s)}(z)$ defined in \re{hatchi}.
At weak coupling, for $y\gg 1$ with $s=2gy$ kept fixed, it is described by the resummed expression \re{tau}. At strong coupling, for $g>y\gg 1$, we use the strong coupling 
expansion of the octagon \re{O-str} with the coefficients given by \re{2nd} and \re{As-beta} for $\beta=1$. In addition, we apply the Borel-Pad\'e method to improve the series 
in $1/g$ (see Ref.~\cite{Belitsky:2020qrm} for details).  The comparison with the numerical values for $y=10$, $\xi=0$ and $\ell=1$ is shown in Figure~\ref{fig:null}.

\subsection*{Large bridge length limit}

\begin{figure}[h!t]
\psfrag{ell}[cc][cc]{$\ell$}\psfrag{logO}[cc][cc]{$\log \mathbb O_\ell$}
\centerline{\insertfig{9}{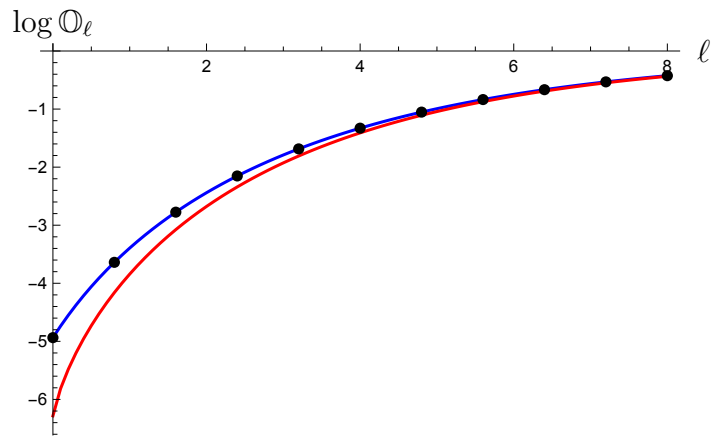}}
\caption{The dependence of the  octagon $\mathbb O_{\ell}(\widehat\chi^{(s)})$ on the bridge length $\ell$
for $g=2$ and $y=\xi=0$.
Comparison of numerical values (black dots) with large $g$ expansion  \re{O-str} (blue curve)
and  large $\ell$ expansion \re{O-bar} (red curve). The two curves 
merge at $\ell \approx 5$, or equivalently $\bar\ell \approx 1.25$.
}\label{fig:large}
\end{figure}

For $\ell=O(g)$ and $g\gg 1$, the leading asymtotic behavior of the octagon is given by \re{O-bar}. As in the previous case, we consider the `simplest' octagon  $\mathbb O_{\ell}(\widehat\chi^{(s)})$ and 
compute it numerically  for various $\ell$'s at some reference value of the coupling constant $g=2$ and kinematical invariants $y=\xi=0$. A comparison of the numerical results with the strong coupling expansion 
 \re{O-str} and the large $\ell$ expansion \re{O-bar} is shown in Figure~\ref{fig:large}.  
 
\subsection*{The Szeg\H{o}-Akhiezer-Kac formula}

Computing the octagon at strong coupling, we applied the conjectured Szeg\H{o}-Akhiezer-Kac formula \re{SAK} for special values of $\beta$ specified in \re{betas}. To test the relation \re{SAK} for arbitrary $\beta$, we choose the symbol of the  Bessel operator to be 
\re{FH} with the regular symbol $b(z)$ being
\begin{align}\label{b}
b(z) = {z^2+1\over z^2} \tanh^2(z/2)\,.
\end{align}
The rational for this choice is that for $\beta=1$ the corresponding function $\widehat\chi(z)$ in \re{FH} coincides with $\widehat\chi^{(s)}(z)$ defined in \re{hatchi} for $y=\xi=0$.

Substitution of \re{b} into \re{psi} yields (see Eqs.~\re{use} and \re{sim1})
\begin{align} 
\widetilde\psi(k) {}&= -{\beta-1\over k}\lr{1-\e^{-k}}- {1\over k} \tanh(k\pi/2)\,.
\end{align}
Taking into account this relation we get from \re{SAK} after some algebra
\begin{align}\notag \label{SAK2}
 \log \det \big( &1-\mathbf B_\ell(\widehat\chi)\big)_{[0,2g]}   =  -( \pi+2\beta-2) g + \frac12  \lr{2\beta\ell + \beta^2}\log g
\\\notag
&  -\frac12 (\beta-1) ^2 \log 2
+2(\beta-1)  \log  {\Gamma
   \left(\frac12+\frac{1}{2\pi }\right) \over \Gamma(\frac12)\Gamma \left(1+\frac{1}{2 \pi }\right) }
   + \frac12\beta\log(2\pi)  
\\&   
  + \ell\log 2  + \log {G(1+\ell)\over G(1+\ell+\beta)} 
+\frac12+\frac12\log \pi +\frac{2 \log 2}{3}-6\log A + O(1/g)\,.
\end{align}
Power suppressed corrections to this relation are given by the last term on the right-hand side of \re{O-str}. It involves the expansion coefficients $A_k$ (with $k\ge 2$) defined in \re{As-beta}. We use \re{Ibeta} and \re{b} to 
find the corresponding profile function as
\begin{align} 
I_n {}&= (-1)^n \left[\beta-1+{2\eta(2n-1)\over \pi^{2n-1}}\right]\,,
\end{align}
where $\eta(x)$ is a Dirichlet eta-function.   
 
\begin{table}[bt!]
\centering
\begin{small}
\begin{tabular}{|ccccccccc|} 
\hline
$(0,0)$ & $(0,\ft14)$ & $(\ft14,0)$ & $(0,\ft12)$ & $(\ft12,0)$ & $(\ft14,\ft14)$ & $(\ft14,\ft12)$ & $(\ft12,\ft14)$ & $(\ft12,\ft12)$ \\[1.2mm] \hline 
-2.21436 & -2.96772 & -2.04107 & -3.66198 & -1.86779 & -2.69201 & -3.31135 & -2.44381 & -3.00826 
\\  
 -2.19493 & -2.95992 & -2.02651 & -3.66198 & -1.86779 & -2.69201 & -3.31385 & -2.45670 & -3.01412 
 \\  
-2.19486 & -2.95993 & -2.02655 & -3.66200 & -1.86792 & -2.69206 & -3.31387 & -2.45676 & -3.01413
\\
\hline
\end{tabular}
\end{small}
\caption{The Szeg\H{o}-Akhiezer-Kac formula \re{SAK2} 
for $g=2$ and different values of the pair $(\ell,\beta)$. 
The entries in the first row are obtained  by neglecting $O(1/g)$ corrections
on the right-hand side of \re{SAK2}, the entries in the second row take into account 
the $O(1/g^6)$ terms. The last row shows the results of numerical evaluation.
For $\ell+\beta=1/2$ the entries in the first and second raws coincide in virtue of the first relation in \re{exa}.
}\label{tab}
\end{table}

To test the relation \re{SAK2}, we evaluated the left-hand side of \re{SAK2} numerically for $g=2$ and different values of $\ell$ and $\beta$, and compared it with the first few terms of the strong coupling expansion.
The results are shown in Table~\ref{tab}.

\section{Quantum curve representation}
\label{AppendixClustering} 
 
We use the relation \re{Y} to verify that 
\begin{align}\label{Y1}
\log(1+Y(\vartheta)) = \log{(1+\e^{-\epsilon_\ell(\vartheta)+y})(1+\e^{-\epsilon_\ell(\vartheta)-y})\over (1-\e^{-\epsilon_\ell(\vartheta)+\xi})(1-\e^{-\epsilon_\ell(\vartheta)-\xi})}\,,
\end{align}
where $ \epsilon_\ell(\vartheta)= \xi\cosh\vartheta+i\bar\ell \sinh\vartheta$. Integrating by parts in \re{i0} we get
\begin{align}\notag\label{quasi0}
\log \mathbb O_\ell^{(s)}  = - {g\over\pi} \int_{-\infty}^\infty {d\vartheta\over \sinh^2(\vartheta-i0 \xi)} \times
\Big[&
{\rm Li_2} (-\e^{-\epsilon_\ell(\vartheta)+y})+ {\rm Li_2} (-\e^{-\epsilon_\ell(\vartheta)-y})
\\
& -  {\rm Li_2} (\e^{-\epsilon_\ell(\vartheta)+\xi})-  {\rm Li_2} (\e^{-\epsilon_\ell(\vartheta)-\xi})
\Big].
\end{align}
Then, we change the integration variable to $u=2g\coth\theta$ and rewrite the above relation as
\begin{align}\notag\label{quasi}
\log \mathbb O_\ell^{(s)}  &= -  {1\over 2\pi}\lr{\int^{-2g}_{-\infty} +\int_{2g}^\infty }{du\over (1-iu\xi \delta)^2}
\\
&\times
\left[
{\rm Li_2} (-\e^{-F_\ell(u)+y})+ {\rm Li_2} (-\e^{-F_\ell(u)-y})-  {\rm Li_2} (\e^{-F_\ell(u)+\xi})-  {\rm Li_2} (\e^{-F_\ell(u)-\xi})
\right],
\end{align}
where $\delta\to 0$ and integration goes over $u^2\ge (2g)^2$. Here the notation  was introduced for
\begin{align}
F_\ell(u) =  {i\ell+\xi u\over \sqrt{u^2-(2g)^2}}  = \ell E _\psi(u) -2i\xi p_\psi(u)\,,
\end{align}
where $E_\psi$ and $p_\psi$ are the (mirror) energy and momentum of excitations that contribute to \re{O-ff}.
The additional factor in \re{quasi} depending on $\delta$ ensures a convergence of the integral at infinity. 
The relation \re{quasi} is similar to  the quantum curve representation of three-point correlation functions discussed in Ref.\ \cite{Jiang:2016ulr}. Repeating 
the same analysis for the octagons $\mathbb O_\pm$ and $\mathbb O_\ell^{(a)}$ we find that they admit similar representations.  
 
\section{Similarity transformation}
\label{app:sim}  
 
In this Appendix, we construct the similarity transformation \re{Omega} for the matrix $H_\pm=\lambda_\pm CK$ which is given by the product of a scalar factor 
$\lambda_\pm$ and two semi-infinite matrices defined in \re{oct-gen} and \re{K-mat}. 

It is convenient to change the integration variable in \re{K-mat} to $x=2g\sqrt{t^2-\xi^2}$ to get another representation for the matrix $K$
\begin{align}\label{K-def1}\notag
& K_{mn} = \frac12 \int_0^\infty dx\,{h_{mn}(x,2g\xi)\over \cos \phi -\cosh \sqrt{x^2/(2g)^2+\xi^2}}\,,
\\[2mm]\notag
& h_{mn}(x,\sigma)= i^{m-n-1} P_{m-n}(\sigma/x) J_{m+\ell} (x) J_{n+\ell}(x)\,,
\\[2mm]
& P_{n}(x)={\lr{x+\sqrt{x^2+1}}^n - \lr{x-\sqrt{x^2+1}}^n  \over 2\sqrt{x^2+1}}\,,
\end{align}
where $\sigma=2g\xi$.
Here $P_n(x)$ satisfies the property $P_{-n}(x) =P_n(-x)=(-1)^{n-1} P_n(x)$. For positive integer $n$, it is given by a polynomial in  $x$ of degree $n-1$ and is related 
to the Chebyshev polynomial of the second kind $P_n(x) =  (-i)^{n-1} U_{n-1}(ix)$. 

Let us introduce notation for the product of two semi-infinite matrices  $\widehat h = C h(x,\sigma)$,
\begin{align}\label{hat-h}
\widehat h_{mn}(x,\sigma) = h_{m+1,n}(x,\sigma) - h_{m-1,n}(x,\sigma) \theta(m-1) 
= \left[\begin{array}{cc}\widehat h_{\rm ee} & \widehat h_{\rm eo} \\ \widehat h_{\rm oe} &  \widehat h_{\rm oo} \end{array}\right]
\,,
\end{align}
where $m,n\ge 0$ and we split $\widehat h_{mn}(x,\sigma)$ into four semi-infinite blocks depending on the parity of $m$ and $n$ (with `e' and `o' referring to even and odd indices, respectively). 
We show below that the $\sigma-$dependence of $\widehat h(x,\sigma)$ can be eliminated by a similarity transformation
\begin{align}\label{h-sym}
\widehat h(x,\sigma) = \Omega^{-1}(\sigma) \, \widehat h(x,0) \,\Omega(\sigma)\,.
\end{align} 
For $\sigma=0$, it follows from \re{K-def1} and \re{hat-h} that $\widehat h_{\rm eo}(x,0)=\widehat h_{\rm oe}(x,0)=0$   and, therefore, the matrix $ \widehat h(x,0)$ takes a block 
diagonal form. Moreover, it is straightforward to verify that nonvanishing entries of $\widehat h_{mn}(x,0)$ obey
\begin{align}
\sum_{k\ge 0} \lr{U_{mk}\widehat h_{2k,2n}(x,0) - \widehat h_{2m+1,2k+1}(x,0) U_{kn}} = 0\,,
\end{align}
where $U_{nm}=(2n+1+\ell)\theta(n-m)$. This relation implies that, for $\sigma=0$, the diagonal blocks of $\widehat h(x,0)$ are related to each other as
\begin{align}
\widehat h_{\rm ee}(x,0) = U^{-1} \, \widehat h_{\rm oo}(x,0) \, U\,.
\end{align}
Then, combining together the relations \re{K-def1}, \re{hat-h} and \re{h-sym}, we find that the matrix $H_\pm=\lambda_\pm CK$ satisfies the relation \re{Omega}.
  
We now observe that the relation \re{h-sym} allows us to determine the matrix $\Omega(\sigma)$ up to a gauge transformation $\Omega(\sigma) \to \omega \,\Omega(\sigma)$ with 
the matrix $\omega$ satisfying $[\omega,   \widehat h(x,0) ]=0$. Using this property, we can fix a gauge by imposing additional conditions on the matrix elements of $\Omega(\sigma)$
\begin{align}\label{mag}\notag
& \sum_{m\ge 0}  \Omega_{2m,2k+1}(\sigma) =\sum_{m\ge 0}  \Omega_{2m+1,2k}(\sigma) = 0\,,
\\
& \sum_{m\ge 0}  \Omega_{2m,2k}(\sigma) = \sum_{m\ge 0}  \Omega_{2m+1,2k+1}(\sigma) = 1\,.
\end{align} 
Expanding both sides of \re{h-sym} at small $\sigma$, it is possible to check that these relations unambiguously fix  the matrix $\Omega(\sigma)$.
Then, we apply \re{h-sym} and \re{mag}  to get for the matrix elements of $ \Omega(\sigma) \widehat h(x,\sigma) = \widehat h(x,0) \,\Omega(\sigma)$
\begin{align} 
\sum_{m,k\ge 0}  \Omega_{2m+1,k} \widehat h_{kn}(x,\sigma) 
 =\sum_{k\ge 0} \widehat h_{2k+1,n}(x,\sigma) =\sum_{m,k\ge 0}\widehat h_{2m+1,2k+1}(x,0)\, \Omega_{2k+1,n} \,,
\end{align}
where in the first relation, we split the sum into the even and odd value of $k$ and used \re{mag} afterwards. In the second relation, we took into account that $\widehat h_{2m+1,k}(x,0)$ 
vanish for even $k$. Replacing $ \widehat h_{2k+1,n}$ with its expression \re{hat-h} and evaluating emerging telescoping sums, we  obtain   
\begin{align}
h_{0 n}  (x,\sigma) = \sum_{k\ge 0} h_{0, 2k+1}  (x,0)\, \Omega_{2k+1,n}(\sigma)\,.
\end{align}
Repeating this analysis for $\sum_{m,k}  \Omega_{2m,k} \widehat h_{kn}(x,\sigma)$, we obtain in a similar manner
\begin{align}
h_{1 n}  (x,\sigma) = \sum_{k\ge 0} h_{1, 2k}  (x,0)\, \Omega_{2k,n}(\sigma)\,.
\end{align}
Replacing $h_{0 n}$ and $h_{1 n}$ with their expressions \re{K-def1} and taking into account that $P_{-n}(x) = (-1)^{n-1} P_n(x)$, we deduce from the last two relations
\begin{align} \notag
& i^{n+1} P_n( \sigma/x) J_{n+\ell}(x) =\sum_{k\ge 0} i^{2k+2}  J_{2k+1+\ell}(x)\, \Omega_{2k+1,n} (\sigma)\,,
\\
& i^{n} P_{n-1}( \sigma/x) J_{n+\ell}(x) = \sum_{k\ge 0} i^{2k}  J_{2k+\ell}(x)\, \Omega_{2k,n} (\sigma)\,.
\end{align}
Finally, we integrate both sides of these relations with $J_{2k+1+\ell}(x)/x$ and $J_{2k+\ell}(x)/x$, respectively, to get $ \Omega_{0,n}=\delta_{n0}$ and
\begin{align}\notag\label{Omega-res}
 \Omega_{2k+1,n} &=
2(2k+1+\ell)(-1)^{k+1}\int_0^\infty {dx\over x} i^{n+1} P_n( \sigma/x) J_{n+\ell}(x)J_{2k+1+\ell}(x) \,,
\\
 \Omega_{2k+2,n} &=2(2k+2+\ell)(-1)^{k+1}\int_0^\infty {dx\over x} i^{n} P_{n-1}( \sigma/x) J_{n+\ell}(x)J_{2k+2+\ell}(x) \,,
\end{align}
where $k,n\ge 0$ and $P_n$ is defined in \re{K-def1}. These relations provide a concise representation of the semi-infinite matrix $\Omega$ entering \re{Omega}. It is straightforward to 
use it to expand $\Omega$ in powers of $\sigma=2g\xi$.

We recall that the relations \re{Omega-res} hold in the gauge \re{mag}. Another representation for the matrix $\Omega$ was recently derived in Ref.~\cite{KP20} using a different 
technique. It differs from \re{Omega-res} by a gauge rotation.
  
\bibliographystyle{JHEP}

\providecommand{\href}[2]{#2}\begingroup\raggedright\endgroup


\end{document}